\documentclass[aps,prb,twocolumn,amsmath,amssymb,nofootinbib,floatfix,eqsecnum]{revtex4-1}
\usepackage{amssymb}

\usepackage{color}
\usepackage{graphicx}

\begin{document}

\title{
Effective field theories for topological insulators 
by functional bosonization
}

\author{
AtMa Chan, 
Taylor L. Hughes, 
Shinsei Ryu
and Eduardo Fradkin}

\affiliation{
 Department of Physics, University of Illinois
 at Urbana-Champaign, 
 1110 West Green St, Urbana IL 61801-3080
             }

\date{\today}

\begin{abstract}
Effective field theories
that describes the dynamics of a conserved
U(1) current in terms of
``hydrodynamic'' degrees of freedom 
of topological phases in condensed matter
are discussed in general dimension $D=d+1$
using the functional bosonization technique.
For non-interacting topological insulators (superconductors)
with a conserved U(1) charge
and characterized by an integer topological invariant
[more specifically, they are topological insulators
in the complex symmetry classes (class A and AIII)
and in the ``primary series'' of topological insulators
in the eight real symmetry classes],
we derive the BF-type topological field theories 
supplemented with the Chern-Simons (when $D$ is odd)
or the $\theta$-term (when $D$ is even). 
For topological insulators characterized by 
a $\mathbb{Z}_2$ topological invariant
(the first and second descendants of the primary series),
their topological field theories are obtained by dimensional reduction. 
Building on this effective field theory description for non-interacting 
topological phases, 
we also discuss,
following the spirit of 
the parton construction of the fractional quantum Hall effect
by Block and Wen, 
the putative ``fractional'' topological insulators
and their possible effective field theories, and use them to determine the physical properties of these non-trivial quantum phases.
\end{abstract}


\maketitle

\tableofcontents 

\section{Introduction}

Topological phases are fully quantum mechanical
states of matter which are not characterized by 
spontaneous breaking of a global symmetry of the quantum mechanical system. 
While gapped in the bulk, quite often
these phases have 
 {\it gapless}
excitations at their boundary, signaling the highly entangled nature
of their ground states.
Since the discovery of the integer quantum Hall effect (IQHE)
and the factional quantum Hall effect (FQHE),\cite{review_QHE}
the list of topological phases in nature has been expanded,
in particular, by 
the recent discovery of time-reversal symmetric topological insulators
in two and three dimensions (2D and 3D)
in systems with strong spin-orbit coupling,
\cite{review_TIa, review_TIb,KaneMele,Roy06, Bernevig05,Moore06,Roy3d,Fu06_3Da,Fu06_3Db,Qi_Taylor_Zhang2008}
and the identification of $^{3}$He-B phase as a topological superconductor
(actually a superfluid).
\cite{3HeB}
Unlike the IQHE,
the topological character of these topological insulators 
and superconductors
({\it i.e.}, the stable gapless edge or surface modes)
is protected by time-reversal symmetry (TRS). 
The presence or absence of a topological distinction among gapped phases
for a given set of symmetries and for given spatial dimensions 
can be studied systematically, and
is summarized in the ``periodic table'' of topological insulators and superconductors for non-interacting fermions.
\cite{Schnyder2008,Kitaev2009,SRFLnewJphys}

One of the defining properties of 
topological insulators is their characteristic 
response to external electromagnetic fields.
A classic example for this is the quantum Hall effect (QHE)
which is characterized by the non-zero, {\it quantized} value
of the Hall conductance.
The three-dimensional time-reversal symmetric topological insulator
is characterized by non-zero magnetoelectric polarizablility,
which, in the presence of time-reversal symmetry, also takes a quantized value.\cite{Qi_Taylor_Zhang2008}
Correspondingly, these responses are described by 
a field theory supplemented with a term of topological origin,
such as the Chern-Simons (CS) term, or the axion term. 
For a wide class of topological insulators and superconductors, 
such topological \emph{response} theories have been studied and carefully classified.
\cite{Qi_Taylor_Zhang2008,Ryu2011,Wang2011,NomuraStreda2011}
When it exists, such {\em topologically protected}
response is a powerful way to
characterize a topological phase,
as it is not limited to non-interacting topological phases.

There is, however, a subtle but profound distinction between a topological phase and a phase with a topologically protected response.
For instance, the IQHE is a phase of a (generally interacting) two-dimensional electron gas in a strong magnetic field whose Hall conductance is precisely defined, and protected, by a topological invariant.\cite{niu-thouless-wu} 
Similarly its edge states are also protected, in this case by their chiral nature. 
However, the fractional quantum Hall states are {\em topological fluids}. 
In addition to having a topologically protected fractional quantum Hall conductance, these fluids are characterized by having non-trivial ground state degeneracies on topologically non-trivial surfaces, excitations  that carry fractional charge and fractional statistics (Abelian or non-Abelian), and by a set of edge states with generally complex properties. The universal properties of such {\em topological phases} are given in terms of effective low-energy hydrodynamic {\rm topological field theories} of these fluids.\cite{wen-niu,wen-zee,Wen1995}

While powerful, the effective field theories of the response of a system to an external local probe should be distinguished from
{\it internal} or {\it hydrodynamic} topological field theories
describing the global properties and the
excitations of topological phases,
as in the example of the FQH fluids.\cite{Wen1995}
Just as fluid 
dynamics is an efficient description of a collection of a macroscopic number
of particles, the dynamics of topological insulators may be well described
in terms of collective degrees of freedom,
rather than relying on 
microscopic electrons (quasi-particles).
In fact, such a picture was developed for the QHE, and we have 
a good understanding of a quantum Hall system as a droplet of electron liquid.
The hydrodynamic topological field theory description of the
quantum Hall droplet 
is given by the Chern-Simons (CS) gauge theory. 
Once such effective description of the low-energy physics is established,
it is likely to be robust against interactions, and has a 
wider range of applicability than the non-interacting microscopic system.
The purpose of this paper is to develop such
a hydrodynamic effective
field theory that is capable of describing
collective excitations in topological insulators in
general dimension.
  For previous studies,
  see, for example,
  Refs.\ \onlinecite{ChoMoore2010,Diamantini2012,Nikolic, Vishwanath2012}.

The CS description of the two-dimensional
QHE can be derived,
for example, from composite particle theories.
In the composite boson theory,\cite{ZhangHanssonKivelson1989}
one first attaches a unit flux 
to each electron to make it a boson. 
For the completely filled Landau Level,
on average
this will 
cancel
the external magnetic field, 
and thus we have a composite boson system 
which is at zero magnetic field, but interacting with
a statistical CS gauge field.
A subsequent duality transformation delivers 
the CS theory which gives us a hydrodynamical
description of the filled Landau level.
Similarly, by using the composite fermion 
picture,\cite{Jain89a,Jain89b} 
the same CS theory can also be derived.\cite{LopezFradkin1991}

The flux attachment concept is very successful to derive the hydrodynamic theories of the Laughlin and Jain states of the FQHE. However, it only works in two dimensions since it is based on the existence of the Chern-Simons gauge theory and a connection with the braiding group. In 2D the excitations (or vortices) can carry the quantum numbers of representations of the braid group which allows for fractional statistics.  
However, the concept of braiding is only meaningful in two spatial dimensions since the braiding of particle worldlines is not topologically stable in other spatial dimensions. 
In addition, for technical reasons, flux attachment is somewhat cumbersome to apply even for to 2D lattice systems (see {\it e.g.}
Ref.\ \onlinecite{FradkinBook}). 
Thus, an alternative procedure to derive effective hydrodynamic field theories of topological phases is desirable.

One alternative approach for deriving  effective field theories is bosonization. This method is closely related to the well-known
boson-fermion correspondence in
 1+1 dimensions which is an exact operator correspondence between a theory of massless (Dirac) fermions and a theory of a massless relativistic scalar field. 
We will show here that
one can use the 
approach known as
``functional bosonization'' to derive effective hydrodynamic field theories for general topological phases both in two and three dimensions.
Although for spatial dimensions $d>1$ the functional bosonization approach is not an exact mapping, 
it is nevertheless 
useful to derive effective low energy theories in massive phases, which is what we are after in this work.
In particular, in this paper, we will apply functional bosonization
to the (3+1)D topological insulator to derive the an effective topological hydrodynamic theory of the form of the BF theory in 3+1 space-time dimensions.

This paper is organized as follows: 
In Sec.\ \ref{sec: Functional bosonization},
we start by applying functional bosonization to
the topological insulators in the primary series of the periodic table of topological insulators.\cite{Schnyder2008,Kitaev2009}
We derive BF-type topological field theories
supplemented with a term of topological origin. 
Next, the physical picture suggested by functional bosonization is described
in
Sec.\ \ref{sec: Physical pictures from the functional bosonization},
making some comparison with flux attachment, and composite particle
theories.
In Sec.\ \ref{sec: Dimensional reduction},
we discuss hydrodynamic topological field theory descriptions
of $\mathbb{Z}_2$ topological insulators,
including the three-dimensional time-reversal symmetric
topological insulators, and the two-dimensional quantum spin Hall effect.
They are obtained from
the BF-type topological field theories of the primary series
by dimensional reduction. {\bf{N.B.}}: We will use $D$ to represent space-time dimensions and $d$ to represent space dimensions. Thus $D=2$ is $1+1$ dimensional space-time.

In Sec.\ \ref{sec: Fractional states},
we explore possibility of topological phases that arise
because of strong interactions, {\it e.g.},
``fractional'' topological insulators.
Starting from the BF theories for non-interacting topological
insulators, and,
following the parton approach pioneered by Blok and Wen\cite{Blok1990a, Blok1990b} in the FQHE
by implementing strong correlations as a constraint
among fields,
we give a derivation of topological field theories
which describe strongly interacting topological phases. 
Finally, we present our conclusions in
Sec.\ \ref{sec: Discussion}.

\section{Functional bosonization}
\label{sec: Functional bosonization}

\subsection{Functional bosonization}

We begin with a summary of functional bosonization. 
Our starting point is 
the fermionic partition function
in $D=d+1$ space-time dimensions, 
\begin{align}
Z[A^{\mathrm{ex}}]
=
\int 
\mathcal{D}
\left[
\bar{\psi},
\psi
\right]
\exp
\left(
{i} K_F[\bar{\psi},\psi,A^{\mathrm{ex}}]
\right),
\end{align}
where 
$K_F[\bar{\psi},\psi,A^{\mathrm{ex}}]$
is the fermionic action describing 
the (topological) insulator in question,
and $A^{\mathrm{ex}}_{\mu}$ is an external 
U(1) gauge field
which couples minimally to the fermion field
$\bar{\psi}, \psi$
as a source (background). 
The correlation functions of
the electrical current operator
can be obtained 
from the generating functional 
$Z[A^{\mathrm{ex}}]$ as 
\begin{align}
&
\langle
j^{\mu_1}(x_1)
j^{\mu_2}(x_2)
\cdots
\rangle
\nonumber \\
&
=
\frac{1}{{i}} \frac{\delta}{\delta A^{\mathrm{ex}}_{\mu_1}(x_1)}
\frac{1}{{i}} \frac{\delta}{\delta A^{\mathrm{ex}}_{\mu_2}(x_2)}
\cdots 
\ln Z[A^{\mathrm{ex}}].
\end{align}

Functional bosonization makes use of the gauge invariance of the action and of the functional integral
under a local U(1) gauge transformation:
for an $a_{\mu}$ which is pure gauge, the partition function is invariant
under the shift $A^{\mathrm{ex}}\to A^{\mathrm{ex}} + a$,
\begin{align}
Z[A^{\mathrm{ex}}+a]= Z[A^{\mathrm{ex}}]. 
\end{align}
Thus, one can represent $Z[A^{\mathrm{ex}}]$ as
\begin{align}
Z[A^{\mathrm{ex}}] = \int 
\mathcal{D}[a]_{\mathrm{pure}}
Z[A^{\mathrm{ex}}+a],
\end{align}
up to some normalization,
where 
$\int
\mathcal{D}[a]_{\mathrm{pure}}
$
is the path integral
over the gauge field $a_{\mu}$
with the condition that it is a pure gauge. Thus, 
in terms of 
the field strength
$f_{\mu\nu}[a]=\partial_{\mu} a_{\nu}-\partial_{\nu} a_{\mu}$, the allowed gauge field configurations are required to obey
$f_{\mu\nu}[a]=0$ for all possible pairs of indices $\mu,\nu$ ($\mu<\nu$).

In the case of  a system of fermions on an open manifold with fixed boundary conditions, such as a disk in two dimensions or a ball in three dimensions,  this procedure does not change the partition function. This is true independently of the dynamics of the fermions (provided it is gauge-invariant). This is also true for a closed topologically trivial manifold (such as a sphere). On the other hand, in the case of fermions on a closed topologically non-trivial  manifold, such as a torus in any dimension, this procedure is equivalent to averaging the partition function over the large gauge transformations on the torus. Thus, this is equivalent to averaging the partition function over ``twisted'' boundary conditions.

The pure gauge condition can be imposed 
by inserting the delta functional 
\begin{align}
\prod_{x}
\prod^{\mu<\nu<\lambda\cdots}_{\mu,\nu,\lambda,\ldots}
\epsilon^{\mu\nu\lambda \cdots\alpha\beta}
\delta
\big([f_{\alpha\beta}[a(x)]
\big)
\end{align}
in the path integral, where
$\epsilon^{\mu\nu\cdots\alpha\beta}$
is the Levi-Civita symbol in $D$ space-time dimensions, 
and 
$\prod^{\mu<\nu<\lambda\cdots}_{\mu,\nu,\lambda,\ldots}$
runs over $D(D-1)/2$ independent directions.
For example,
when $D=3$ and $D=4$,
the delta functional is given by
$
\prod_x \prod^3_{\mu=1}
\epsilon^{\mu\nu\lambda}
\delta(f_{\nu\lambda}[a])$, 
and
$\prod_x \prod^{\mu<\nu}_{\mu,\nu}
\epsilon^{\mu\nu\lambda\sigma}
\delta(f_{\lambda\sigma}[a])
$,
respectively. 
The delta functional can be exponentiated 
by introducing
an auxiliary 
rank $(D-2)$ tensor field
$b_{\mu\nu\ldots}$ as
\begin{align}
&Z[A^{\mathrm{ex}}] 
= 
\int \mathcal{D}[a,b]
Z[A^{\mathrm{ex}}+a]
\nonumber \\
&\qquad 
\times 
\exp \Big(-
\frac{{i} }{2}
\int d^Dx\,
b_{\mu\nu\cdots}
\epsilon^{\mu\nu\cdots\alpha\beta}
f_{\alpha\beta}[a]\Big).
\end{align}
Using the invariance of the integration measure under a shift of the gauge fields, $a\to a-A^{\mathrm{ex}}$, we can write the partition function $Z[A^{\rm ex}]$ as
\begin{align}
&
Z[A^{\mathrm{ex}}] 
= 
\int \mathcal{D}[a,b]
Z[a]
\nonumber \\
&\quad 
\times 
\exp\Big(-
\frac{{i} }{2}
\int d^Dx\,
b_{\mu\nu\cdots}
\epsilon^{\mu\nu\cdots\alpha\beta}
\left(
f_{\alpha\beta}[a]
-
f_{\alpha\beta}[A^{\mathrm{ex}}]
\right)\Big).
\label{general}
\end{align}
From the partition function (\ref{general}), 
the current correlation functions can be computed as
the correlation functions of the tensor field $b_{\mu\nu\cdots}$,
\begin{align}
&
\langle 
j^{\mu_1}(x_1)  j^{\mu_2}(x_2)\cdots
\rangle
\nonumber\\
&
\quad =
\langle
 \epsilon^{\mu_1\nu_1\lambda_1\cdots}\partial_{\nu_1} b_{\lambda_1\cdots} (x_1)
 \epsilon^{\mu_2 \nu_2\lambda_2\cdots}\partial_{\nu_2} b_{\lambda_2\cdots} (x_2)
\cdots  
\rangle
,
\end{align}
suggesting the correspondence (bosonization rule)
\begin{align}
j^{\mu}(x)
\quad
\Leftrightarrow 
\quad
\epsilon^{\mu\nu\lambda\rho\cdots}\partial_{\nu} b_{\lambda\rho\cdots} (x). 
\label{bosonization-rule}
\end{align}
where the $(D-2)$-form tensor field $b_{\mu\nu\cdots}$ was
introduced as a Lagrange multiplier field. 
This system has in fact a (local) gauge symmetry as its partition function is invariant under
\begin{align}
b_{\mu\nu\cdots}
&\to
b_{\mu\nu\cdots} + 
\partial_{\{ \mu} \xi_{\nu\lambda\cdots \}}
\end{align}
where the symbol $\{\cdots\}$ fully antisymmetrizes the indices. 
Observe also that
in Eq.\ (\ref{general})
the field strength for
$b_{\mu\nu\cdots}$, 
which is the $(D-1)$-form field $h$ defined by $h:=db$
(in the differential form notation),
does not appear. 

A consequence of the bosonization rule Eq.\ \eqref{bosonization-rule}
and of the form of the partition function, Eq.\ \eqref{general}, 
is that local magnetic fluxes 
(or, in general dimension, magnetic flux ``tubes'') couple to the tensor field $b$. In addition, in a system with an energy gap, the worldlines of fermionic excitations 
({\it i.e.} the quasiparticle currents) are minimally coupled to the gauge field $a_\mu$.

To complete the bosonization mapping,
we need to evaluate the fermionic path integral $Z[a]$. 
Our discussion so far is quite general
and applicable to gapped/gapless, interacting/non-interacting systems in any dimension.
 However, for a theory of massless fermions (free or interacting)
the fermionic path integral $Z[a]$ 
can only be evaluated exactly in $D=1+1$ dimensions.
In general space-time dimension $D$ the fermion path integral $Z[a]$ is a non-local (but gauge invariant) functional of the gauge field. On the other hand, if the band gap (mass) is finite, 
an approximate form of the fermionic path integral $Z[a]$
can be obtained using the inverse mass expansion. Below, 
we will discuss first the case of non-interacting (topological) insulators and we will later extend these results to the interacting cases.
The path integral calculation is most conveniently described 
by choosing 
the Dirac representative of topological insulators
as a microscopic model. 
The resulting  bosonized action is then a sum of local gauge-invariant operators. In particular the parts of
it which are topological,
do not depend on microscopic details.
These topological terms are always marginal and dominate the low-energy physics for $d<3$,
and they also only depend on universal properties of the system. In particular they contain the information on the 
topological invariants of the microscopic model. 

In the next subsection we discuss some technical and subtle points about functional bosonization that might be skipped during a first reading. After this discussion we move on to discuss what this approach predicts in various dimensions $D$ using several examples.

\subsection{Generalities} 
\label{subsec generalities}

The functional bosonization technique is rooted in
the well-known fermi-bose equivalence in $D=1+1$ dimensions. 
\cite{MattisLieb1965,LutherPeschel1974,LutherEmery1974,Coleman1975,Mandelstam1975,Haldane1981,Gamboa1984,Polyakov-Wiegmann}
(For reviews on bosonization see
Refs.\ \onlinecite{FradkinBook, Stone1994,Gogolin1998}).
It is known that the density-density commutator of a $D=1+1$-dimensional 
chiral fermion is anomalous in the sense that
it develops the so-called Schwinger term,
while one would naively expect 
 (for a system with a relativistic spectrum)
that the charge and current density operators at different
positions (momenta) commute with each other
at equal times. This is due to the underlying chiral anomaly.
This anomalous commutator allows 
one to represent the fermion density operator
in terms of a boson field.
Indeed, if we represent the (normal-ordered) operators
for the fermion charge density $j_0$ and current density $j_1$
by the two-vector $j_\mu=(j_0,j_1)$,
then the anomalous commutator
$[j_0(x),j_1(y)]=(i/\pi) \partial_x \delta(x-y)$
implies that the current can be represented in terms of a scalar (Bose) field $\phi$, {\it i.e.} $j_\mu= (1/\sqrt{\pi}) \epsilon_{\mu \nu} \partial^\nu \phi$, which is consistent with the requirement of 
 local conservation of the charge current, $\partial_\mu j^\mu=0$. In the Abelian case, the resulting effective field theory for the bose field $\phi$ turns out to be free (up to a finite renormalization of the Luttinger parameter and the speed of excitations due to interaction effects). In other words, the dynamics of the fermionic system (interacting or not) is fully represented in terms of the dynamics of its conserved currents. In this sense, in one space dimension, bosonization yields an exact hydrodynamic representation of the system.
In fact, the full excitation spectra of both systems are identical and so are their partition functions. This exact fermi-bose equivalence is a consequence of the kinematic restrictions of one-dimensional motion for systems with a relativistic (linear) dispersion.  Using these identities
one can establish a one-to-one mapping between seemingly different systems such as
the $D=2$-dimensional massive Thirring model
and
the $D=2$-dimensional sine-Gordon model.\cite{Coleman1975,Mandelstam1975}
Such a correspondence also extends to 
$D=2$-dimensional theories with non-Abelian symmetries.\cite{Witten1984}

In $1+1$ space-time dimensions these bosonization operator identities   
can also be formulated in terms of the
functional integral language. In this (functional) bosonization approach the fermion determinant of a Dirac operator coupled to gauge fields
is computed by means of a local chiral transformation.
At the classical level ({\it i.e.} in the Lagrangian) in $1+1$ dimensions it is possible to (formally) cancel the coupling of the fermi field to a gauge field $A_\mu(x)$ by  means of a local chiral transformation, which acts of the Dirac fermion as $\psi(x) \to \exp[i \phi(x) \gamma_5] \psi(x)$, where $\phi(x)$ is an
arbitrary smooth function of
space-time coordinates $x$
and $\gamma_5$ is a suitable Pauli matrix. In a gauge-invariant theory the chiral anomaly implies that the integration measure of the fermion path integral is not invariant under the chiral transformation. As a result there is  a non-trivial Jacobian associated with the chiral transformation.\cite{Fujikawa} This Jacobian is computable in terms of the fermion determinant which leads to the bosonized form of the effective action theory which is a local functional of the bose field $\phi(x)$.
This approach has been applied with great success to several $1+1$-dimensional theories of massless fermions.\cite{Gamboa1984} 
In massive theories, the fermi-bose mapping  relates the two descriptions as identities valid order by order in a perturbation theory in powers of the fermion mass term.\cite{Coleman1975,Mandelstam1975,Naon1985}
It also yields an alternative derivation of non-Abelian bosonization yielding the Wess-Zumino-Witten model\cite{Witten1984}  by means of an exact computation of the fermion determinant.\cite{Polyakov-Wiegmann}

For $d>1$ it is not possible, in general, to find an exact mapping in the form of operator identities,  between a \emph{local} theory of fermions and a \emph{local} theory of bosons. In practice, for $d>1$ bosonization reduces to finding an effective field theory in terms of bose fields. Physically this is largely due to the fact that the kinematic restrictions of one dimension do not exist for $d>1$ and,  as a result, the spectrum of a local fermionic theory, interacting or not, is not equivalent to the spectrum of a local bosonic theory.\cite{comment-tomography} The resulting effective field theory is local only in the case of theories with a finite gap in the fermionic spectrum and for energies much smaller than the gap. The prime example of such a theory is the hydrodynamic field theory of the FQHE. 

For the case of functional bosonization, the problem is that the resulting fermion determinant for massless fermions coupled to gauge fields is a gauge-invariant but non-local functional of the gauge field and, contrary to the case of $D=2$, cannot be computed in closed form. However in the case of massive fermions the contribution of the fermion determinant to the effective action of the gauge fields can be expressed in terms of local gauge-invariant terms (for manifolds without boundary) with coefficients that involve powers of the inverse of the mass, {\it i.e.} a gradient expansion. This approach naturally is only meaningful for massive theories. In the case of massive Dirac fermions in $D=2+1$ dimensions it has been shown\cite{Fradkin1995,Schaposnik1995,Barci2000,BarciOxman2000,Shizuya2001,Schaposnik2001}
that the correlation functions of the conserved fermionic currents are the same as the correlation functions of a {\em dual} topological Chern-Simons gauge theory  in the low energy regime (only!). A mapping of the correlators of other fermion bilinears can also be determined but, again, only in the low-energy and long wavelength regime.

Below,
we will show that functional bosonization can be used to derive hydrodynamic theories of topological insulators in $d=1,\ldots,4$. 
Specifically, we give the functional bosonization results
for non-interacting topological insulators in 
$D=d+1$ with $d=1,\ldots, 4$
which crucially conserve electromagnetic U(1) charge.
These topological phases belong to 
``the complex symmetry classes''
and to
``the primary series'' 
of the eight real symmetry classes
in the periodic table;
by ``complex symmetry classes'' we refer to 
symmetry classes A and AIII,  
while
the ``primary series'' of topological insulators (superconductors) 
is located on the diagonal of the periodic table,
and 
are characterized by 
an integer $\mathbb{Z}$ topological invariant.

After this we will discuss 
field theory descriptions of 
non-interacting topological phases 
characterized by $\mathbb{Z}_2$ invariants.
These theories can be obtained from ``the primary series'' 
by dimensional reduction and can be divided into two different classes:
first and second descendants. 
These phases include
the time-reversal symmetric topological insulators 
in $D=3+1$ (a first descendant) and $D=2+1$
(the quantum spin Hall effect, a second descendant) dimensions. 
It is also possible to extend our discussion
to topological phases (topological superconductors)
which conserve non-Abelian currents
such as a spin SU(2) current. 
\cite{diamond09}
Such non-Abelian functional bosonization
is discussed in Appendix \ref{non-Abelian functional bosonization}.

\subsection{Functional Bosonization Examples} 
\label{subsec examples} 
\subsubsection{$D=1+1$ (AIII or BDI)}

There are no non-trivial topological insulator states in $D=1+1$ dimensions 
if we do not require a protected symmetry. 
That is, 
topological insulators (in fact, topological phases in general) in $D=1+1$
must be  symmetry-protected ones.
In the following, we consider band insulators 
with sublattice (chiral) symmetry. 
In the Altland-Zirnbauer classification,\cite{Altland-Zirnbauer} 
they belong to symmetry class AIII. 
In $D=1+1$, 
topological insulators in symmetry class AIII
are characterized by an integer-valued topological invariant,
the winding number $\nu$. 
A canonical example for such systems is polyacetylene.\cite{Jackiw-Schrieffer}

In addition to the sublattice symmetry,
we can further impose time-reversal symmetry
which squares to be either $+1$ or $-1$ 
(corresponding to symmetry class BDI or CII). 
In either case, 
band insulators in $D=1+1$ dimensions
in these symmetry classes 
are 
characterized by an integral topological invariant $\nu$,
in much the same way as symmetry class AIII in $D=1+1$. 
Below, we will consider symmetry class BDI in $D=1+1$,
which is in the primary series in the periodic table.  

The symmetry class BDI can be realized either in terms of 
complex (Dirac) or real (Majorana) fermions. 
\cite{FidkowskiKitaev2010,FidkowskiKitaev2011,Ryu2011}
In the latter case, instead of imposing a sublattice symmetry, 
we would impose a reality condition (originating from the fact that we are dealing
with Majorana fermions) combined with time-reversal symmetry. The reality condition is equivalent to a charge-conjugation symmetry and when combined with time-reversal
plays a role similar to sublattice symmetry.
Our discussion below, however, will focus on the case of complex
fermions, as functional bosonization takes advantage of the
presence of a U(1) global symmetry.

To summarize, we consider (topological) band insulators
with U(1) global symmetry (particle number conservation) 
in symmetry class AIII or BDI in $D=1+1$ dimensions. 
For these cases,
$Z[a]$ in
Eq.\ (\ref{general})
can be evaluated  as
\begin{align}
\ln Z[a]
&=
\frac{{i} \theta}{2\pi} 
\int d^2x\, 
\epsilon^{\mu\nu}
\partial_{\mu} a_{\nu}
+
\cdots. 
\label{fermion det in 1+1}
\end{align}
Here, the angle $\theta$ is given in terms of 
the bulk topological invariant, the winding number  $\nu$,
as
$\theta = \nu \pi$ mod $2\pi$. 
In Eq.\ (\ref{fermion det in 1+1}), 
terms with more fields and derivatives
such as Maxwell terms
are suppressed as they are more irrelevant
[indicated by $\cdots$ in Eq.\ (\ref{fermion det in 1+1})].

Thus, 
the bosonized partition function is given by
\begin{equation}
Z[A^{\mathrm{ex}}] 
= \int 
\mathcal{D}[a, b]
\exp {i} \int d^Dx \mathcal{L}
\end{equation}
with the effective low-energy Lagrangian density
\begin{align}
&\mathcal{L} =
-b \epsilon^{\mu\nu} \partial_{\mu} 
(a_{\nu}-A^{\mathrm{ex}}_{\nu})
+
\frac{\theta}{2 \pi} 
\epsilon^{\mu\nu}
\partial_{\mu} a_{\nu}
+
\cdots. 
\end{align}
In summary, the effective low-energy theory of massive fermions in $1+1$ dimensions is a BF theory (the first term) and a topological term whose coupling constant is the topological invariant $\theta=\pi \nu$ where $\nu$ is the winding number. Notice that in $D=1+1$ dimensions the BF theory involves a vector (gauge) field $a_\mu$ and a scalar field $b$.

\subsubsection{$D=2+1$ (A or D)}

In $D=2+1$, we discuss the bosonization of 
topological insulators in symmetry classes A and D.
They are topological insulators belonging to 
the primary series in the terminology introduced above. 
These topological insulators are 
characterized by an integer topological invariant,
$\mathsf{Ch}$, 
the first Chern number, 
which is nothing but the Hall conductivity $\sigma_{xy}$
(in units of $e^2/h$). 
The symmetry class A is defined as a set of fermionic quadratic
Hamiltonians which possess no discrete symmetry, and 
the canonical example of the topological insulator in this class
is the integer QHE. 
The symmetry class D is obtained, from symmetry class A, 
by imposing additional particle-hole symmetry while
keeping the electromagnetic U(1) symmetry. 
\cite{Qi_Taylor_Zhang2008}
This should be distinguished from the superconducting
realization of symmetry class D, where there is no 
conserved U(1) charge whereas the particle-hole ``symmetry''
simply reflects the fact that fermionic Bogoliubov 
quasiparticles must satisfy a reality constraint.

For these cases,
$Z[a]$ can be computed in the low-energy limit
as
\begin{align}
\ln Z[a]
&
=
\frac{{i} \mathsf{Ch}}{4\pi} 
\int d^3x\, 
\epsilon^{\mu\nu\lambda}
a_{\mu}\partial_{\nu} a_{\lambda}
+
\cdots, 
\end{align}
where $\mathsf{Ch}$ is the integer-valued  bulk topological invariant,
the first Chern-number, of the topological insulator. 
As in the case of $D=2$, terms which are less relevant 
compared the Chern-Simons term, including the Maxwell term,
are suppressed by powers of the inverse of the mass (the energy gap).
Thus, 
\begin{equation}Z[A^{\mathrm{ex}}] 
= \int 
\mathcal{D}[a, b]
\exp {i} \int d^Dx \mathcal{L},
\end{equation}
with the effective low-energy Lagrangian density
\begin{align}
\mathcal{L} =
-b_{\mu} \epsilon^{\mu\nu\lambda} \partial_{\nu} 
(a_{\lambda}-A^{\mathrm{ex}}_{\lambda})
+
\frac{\mathsf{Ch}}{4\pi} 
\epsilon^{\mu\nu\lambda}
a_{\mu}\partial_{\nu} a_{\lambda}. 
\label{bosonized action D=3}
\end{align}
Thus, here too we obtain a BF theory (this time in $2+1$ dimensions) and a topological invariant term, the Chern-Simons action, whose coupling constant (or ``level'')  is the (topological invariant) Chern number of the microscopic band. In $D=2+1$ the BF term relates two gauge fields, $b_\mu$ and $a_\mu$. This effective action has a formal similarity  with the hydrodynamic theory of the FQHE. However in the hydrodynamic theory of the FQHE  the two gauge fields are  the hydrodynamic field $b_\mu$, and the statistical gauge field $a_\mu$, with the important difference that the Chern-Simons term affects the hydrodynamic field.
\cite{Wen1995,Lopez1999} In the present case the bosonized action, 
Eq.\ \eqref{bosonized action D=3},
represents an integer QHE with a quantized Hall conductance $\sigma_{xy}=({e^2}/{h}) \mathsf{Ch}$. Thus, this is a theory of the quantized anomalous Hall effect.

\subsubsection{$D=3+1$ (AIII or DIII classes)}

In $D=3+1$, 
topological insulators in symmetry class AIII 
are characterized by an integral topological invariant, the three-dimensional integer-valued winding number $\nu$,
and are protected by a chiral symmetry. 
An  example of topological insulators in
this class can be found in Ref.\ \onlinecite{Hosur2009} which discusses a lattice tight-binding model description.
This topological insulator is somewhat analogous to
the time-reversal symmetric topological insulator 
in symmetry class AII, in that it supports a Dirac fermion 
surface state, 
and has a non-trivial
axion-electrodynamics response to the external electromagnetic field.
The difference is, however, that the latter is characterized by
a $\mathbb{Z}_2$ topological invariant, rather than an integer
topological invariant. We will discuss the bosonization of 
the time-reversal symmetric topological insulator 
in symmetry class AII in a later section. 

The topological insulator in symmetry class DIII is 
similar to the topological insulator in 
symmetry class AIII above but with the requirement of an additional
particle-hole or time-reversal symmetry (which lead to equivalent results since these symmetries, when combined with the sublattice symmetry, give the third). 
As in the cases of $D=2$ and $D=3$ above, 
this realization of symmetry class DIII in terms of 
complex fermions ({\it i.e.,} with conserved U(1) charge)
should be distinguished
from the superconducting realization of symmetry class DIII.

For these cases,
$Z[a]$ can be computed, again in the low-energy limit,  
as (see, for example, Ref.\ \onlinecite{Hosur2009} for 
calculations in terms of the Dirac representative), 
\begin{align}
\ln Z[a]
&
=
\frac{{i} \theta}{8 \pi^2} 
\int d^4x\, 
\epsilon^{\mu\nu\lambda\rho}
\partial_{\mu} a_{\nu}\partial_{\lambda} a_{\rho}
+
\cdots. 
\end{align}
Here the angle $\theta$ is related to the 
winding number as
$\theta = \nu \pi$ mod $2\pi$. 
\cite{SRFLnewJphys}
Thus, 
\begin{equation}
Z[A^{\mathrm{ex}}] 
= \int 
\mathcal{D}[a, b]
\exp {i} \int d^D x \mathcal{L}
\end{equation}
with the effective Lagrangian
\begin{align}
&\mathcal{L}
=
-b_{\mu\nu} 
\epsilon^{\mu\nu\lambda\rho} 
\partial_{\lambda} (a_{\rho}-A^{\mathrm{ex}}_{\rho})
+
\frac{\theta}{8\pi^2} 
\epsilon^{\mu\nu\lambda\rho}
\partial_{\mu} a_{\nu}\partial_{\lambda} a_{\rho}
\nonumber \\
&\qquad 
-
\frac{1}{4\pi^2 g^2}
	f_{\mu\nu}f^{\mu\nu}
+
\cdots. 
\label{bosonized action D=4}
\end{align}
In the last line, we have written down the Maxwell term explicitly since it is also marginal in $D=4$.

\subsubsection{$D=4+1$ (A or AII classes)}

Finally, we discuss the bosonization of 
the topological insulators in $D=4+1$. 
In fact, the bosonization rule applies to 
any dimension. 
One of our motivations to study the case of $D=4+1$
is that the lower dimensional topological insulators,
in particular, the time-reversal symmetric 
topological insulators in $D=3+1$ and $D=2+1$,
are the first and second descendants of the $D=4+1$ topological insulator in class AII. 
\cite{Qi_Taylor_Zhang2008, SRFLnewJphys}
In $D=4+1$, 
topological insulators in symmetry class A and 
AII are characterized by an integer-valued topological invariant,
the second Chern number $\mathsf{Ch}_2$.

For these cases,
$Z[a]$ can once again be computed as
\begin{align}
\ln Z[a]
&
= 
\frac{{i} \mathsf{Ch}_2}{24\pi^2}
\int d^5 x\, 
\epsilon^{\mu\nu\lambda \rho\sigma}
a_{\mu} 
\partial_{\nu}
a_{\lambda} 
\partial_{\rho}
a_{\sigma}
\nonumber\\ &-
\frac{1}{4\pi g^2}
f_{\mu\nu}f^{\mu\nu}
+
\cdots, 
\end{align}
where the integer $\mathsf{Ch}_2$
is the second Chern number of the topological insulator. 
Thus, 
\begin{equation}
Z[A^{\mathrm{ex}}] 
= \int 
\mathcal{D}[a, b]
\exp {i} \int d^D x \mathcal{L}
\end{equation}
where the effective Lagrangian is now given by
\begin{align}
\mathcal{L}
&=
-b_{\mu\nu\lambda} 
\epsilon^{\mu\nu\lambda\rho\sigma} \partial_{\rho} 
(a_{\sigma}-A^{\mathrm{ex}}_{\sigma})
\nonumber \\
&\quad 
+
\frac{\mathsf{Ch}_2}{24\pi^2}
\epsilon^{\mu\nu\lambda \rho\sigma}
a_{\mu} 
\partial_{\nu}
a_{\lambda} 
\partial_{\rho}
a_{\sigma}
+
\cdots 
\end{align}
where we have dropped the Maxwell term in the last equation.

\subsubsection{Interactions}

We have not included interactions so far, 
but perturbative effects of weak interactions
can easily be taken into account. 
As an example, consider the local current-current interaction (which is the analog of the Luttinger-Thirring interaction in $D=1+1$ dimensions)
\begin{align}
 -u \int d^Dx\, j^{\mu} j_{\mu},
\end{align}
where $u$ is the coupling constant.
Because of the bosonization rule, 
this leads to a term of the form
\begin{align}
 -u
\int d^Dx\,
(\epsilon^{\mu\nu\lambda\rho \cdots} \partial_{\nu} b_{\lambda \rho \cdots })
(\epsilon_{\mu \alpha\beta\cdots} 
\partial^{\alpha} b^{\beta \gamma \cdots }).
\end{align}
That is, the interactions are represented by a 
Maxwell-like-term 
for the $b$ field.

\section{Physical pictures from functional bosonization}
\label{sec: Physical pictures from the functional bosonization}

In this section, we discuss the physical picture that
functional bosonization provides 
in three different situations
with increasing complexity: 
an example of a topologically trivial insulator,
a time-reversal breaking topological insulator (the QHE or quantum anomalous Hall effect) 
in $D=2+1$ dimensions,
and
a time-reversal invariant topological insulator in $D=3+1$ dimensions.

\subsection{Topologically trivial insulators}
\label{subsubsec topologically trivial insulators}

We start our discussion from topologically trivial insulators
and we will make a few comments. We will discuss:
(i)  the normalization of the gauge fields when
the BF theory description is considered on a compact spatial manifold
(ii) functional bosonization and the dual picture of insulators, and
(iii) the presence/absence of fermionic excitations in the low-energy spectrum. 
In fact, these comments are not limited to topologically trivial insulators,
and we will continue our discussion of them
for non-trivial topological insulators in the later
sections.

(i)
Functional bosonization gives us a bosonized description
of non-topological band insulators
in terms of the BF theory 
{\it without} a CS or $\theta$ term,
{\it i.e.}, the pure BF theory.
In $D=2+1$ dimensions,
the bosonized Lagrangian of 
a non-topological band insulator
is given by Eq.\ (\ref{bosonized action D=3})
with vanishing Chern number, $\mathsf{Ch}=0$.
On a compact spatial manifold, such as a torus,
we should be careful of the normalization of the gauge fields
which must be done
relative to their compactification condition.
On the torus,
where the spatial coordinates $(x_1,x_2)$ are identified
periodically as $x_i \equiv x_i+L_i$, 
the gauge fields $\alpha^i_{j}$
($\alpha^1_{\mu}\equiv b_{\mu}$
and
$\alpha^2_{\mu}\equiv a_{\mu}$),
in the $\alpha^i_0=0$ gauge, 
are normalized such that 
they are global gauge-transform equivalent 
to
$
\alpha^i_{j} + 2\pi n_j/L_j
$
where $n_{j=1,2}$ are integers. 
When properly normalized,
the BF Lagrangian is given by
\begin{align}
\mathcal{L} 
&=
-
\frac{2\mathsf{k}}{4\pi} 
b_{\mu} \epsilon^{\mu\nu\lambda} \partial_{\nu} 
(a_{\lambda}-A^{\mathrm{ex}}_{\lambda}), 
\label{bosonized action D=3 no Ch}
\end{align}
with $\mathsf{k}=1$.
With this normalization,
the ground state degeneracy is one
(\emph{i.e.} no ground state degeneracy).

(ii)
The BF theory description of (non-topological) band insulators
resembles the dual description of BCS superconductors. 
\cite{Balachandran1993,
Bergeron1995,
  Hansson04}
The bosonized theory of the trivial-band insulator
from functional bosonization
is naturally related to 
the dual theory of (band/Mott) insulators.
\cite{LeeKivelson2003,Shindou07}
Below, let us take a closer look at this by focusing on $D=2+1$,
although a similar discussion applies for $D\neq 2+1$.

In (2+1)-D superconductors, 
a defect (vortex) of the condensate 
$\Delta(\vec{r})$
traps a magnetic field
$\nabla \times 
\boldsymbol{A}_{\mathrm{EM}}(\vec{r})$;
the phase $\phi(\vec{r})$ of the superconducting 
order parameter 
$\Delta(\vec{r})
=|\Delta(\vec{r})|
\exp {i}\phi(\vec{r})$
is related to the magnetic field
as
$
(2\pi)^{-1}\Phi_0 \oint_{\partial S} 
{\nabla} \phi
\cdot d\boldsymbol{l}
=
\int_{S}  
({\nabla} \times \boldsymbol{A}_{\mathrm{EM}})
\cdot d\boldsymbol{S},
$
where 
$\Phi_0$ is the flux quantum, 
and if we reinstate $h$, $c$, and $e$, 
$\Phi_0=h c/(2e)$; 
the magnetic flux is localized in the region $S$.

In the dual description of an insulator, 
we view the charge density modulation 
$\delta \rho(\vec{r})$ caused, perhaps, by external doping,
as a point-like ``defect''.  
As in the (2+1)d superconductor, 
we postulate that such a defect traps a
fictitious magnetic field
$\nabla \times \boldsymbol{a}$
\begin{align}
\int_S 
\delta \rho
\, {d}S
&=
e \int_S
(\nabla \times \boldsymbol{a})\cdot {d}\boldsymbol{S},
\label{rho vs flux}
\end{align}
where $a_{\mu}$ is a fictitious gauge field. 
As will become clear,
this gauge field is the 
same gauge field appearing in Eq.\ (\ref{bosonized action D=3 no Ch}).
As the local charge density modulation
is viewed as a vortex, 
we can postulate the existence of an order parameter
$\chi$.
The phase of $\chi$,
$\chi=|\chi|\exp {i}\Theta$, 
is related to the charge density as
\begin{align}
e\oint_{\partial S} \nabla \Theta \cdot 
{d}\boldsymbol{l}
=
\int_S \delta \rho
{d}S.
\label{Theta vs rho}
\end{align}

As in the dual theory of superconductors,
the above description in terms $a_{\mu}$ and $\chi$
can be dualized (in the condensed phase)
and can instead be rewritten in terms of two gauge
fields $a_{\mu}$ and $b_{\mu}$, which are the gauge fields
appearing in Eq.\ (\ref{bosonized action D=3 no Ch}).
Upon dualization,
$b_{\mu}$ 
couples minimally to the vortex current of $\chi$, 
$j^{\mu}_{V}=\epsilon^{\mu\nu\lambda} \partial_{\nu} \partial_{\lambda} \Theta$.
This is in fact consistent with
the bosonization rule
$j^{\mu} \propto  \epsilon^{\mu\nu\lambda}\partial_{\nu} b_{\lambda}$.

(iii)
The bosonic descriptions (in terms of $a_{\mu}$ and $\chi$,
or in terms of $a_{\mu}$ and $b_{\mu}$) do not contain
any fermionic excitations even though
we started from a gapped fermionic theory (a band insulator). 
Drawing again an analogy to BCS superconductors,
this resembles to 
the fact that, under duality, a superconductor 
in a uniform field is equivalent to the insulating phase
of the same bosonic theory at uniform charge density.
\cite{Peskin1978,Dasgupta1981,Fisher1989}
In the latter theory, 
the charges are gapped Cooper pairs,
not fermions.
For this reason the bosonic theory
only describes arrays of Josephson junctions
in which the fermionic excitations do no exist
(they are bound on each superconducting grain).
The connection between the gapped fermionic theory and the bosonized descriptions
is that they share the same topological
limit in the form of a BF theory.
As we will see in the next section, the bosonized description of a 
$D=2+1$ dimensional topological insulator is different and supports
fermionic excitations (electrons) 
due to the presence of the Chern-Simons term.

\subsection{Topological insulator in $D=2+1$}

Let us now move on to a more non-trivial discussion of a topological insulator. 
The bosonized Lagrangian of 
the $D=2+1$ dimensional topological insulator
(in symmetry class A or D) 
is given by
Eq.\ (\ref{bosonized action D=3}):
\begin{align}
\mathcal{L} 
&=
-
\frac{2\mathsf{k}}{4\pi} 
b_{\mu} \epsilon^{\mu\nu\lambda} \partial_{\nu} 
(a_{\lambda}-A^{\mathrm{ex}}_{\lambda})
+
\frac{\mathsf{Ch}}{4\pi} 
\epsilon^{\mu\nu\lambda}
a_{\mu}\partial_{\nu} a_{\lambda}
\nonumber \\
&=
\frac{K_{ij}}{4\pi}
\alpha^i_{\mu} \epsilon^{\mu\nu\lambda} \partial_{\nu} 
\alpha^j_{\lambda}
+ 
\frac{\mathsf{k}}{2\pi} 
b_{\mu} \epsilon^{\mu\nu\lambda} \partial_{\nu} 
A^{\mathrm{ex}}_{\lambda},
\label{bf-cs theory}
\end{align}
where 
we integrated by parts once, and introduced 
the following notation:
$(\alpha^1_{\mu},\alpha^2_{\mu})=(b_{\mu},a_{\mu})$ and 
\begin{align}
K
=
\left(
\begin{array}{cc}
0 & -\mathsf{k} \\
-\mathsf{k} & \mathsf{Ch}
\end{array}
\right).
\end{align}
We have normalized $b_{\mu}$ as before (assuming we are on a compact manifold such as a torus)
and $\mathsf{k}=1$.

This effective theory also makes sense for $\mathsf{k}\neq 1$.
However in this case this effective hydrodynamic theory cannot be
derived from a free-fermion system, even if it is a Chern insulator.
In fact, this effective field theory describes a topological fluid.
This can be checked by noting that it describes a system with a
non-trivial ground state degeneracy $\mathsf{k}^g$, where $g$
is the number of handles of the surface.
In other terms, a theory with $\mathsf{k}\neq 1$ describes a
fractionalized topological insulator,
{\it e.g.} a fractional quantum Hall fluid,
as we will see in Section \ref{sec: Fractional states}.

The Lagrangian (\ref{bf-cs theory})
is the hydrodynamic effective field theory
for $D=2+1$ dimensional topological insulators.
Some examples include the QHE realized in
a two-dimensional electron gas with Landau levels produced by a uniform magnetic field,
as well as
the Chern-insulator
{\it i.e.}, lattice fermion systems
with non-zero Chern number but 
without uniform magnetic field, such as the
Haldane model\cite{Haldane1983}).
Note that we have arrived at
Eq.\ (\ref{bosonized action D=3})
without using composite particle theories
(see below for more comments); and that
for the Chern-insulator,
the flux attachment transformation is highly non-trivial
but not impossible.
\cite{Fradkin1989,Murthy2011,Murthy2012}

It is readily seen that the bosonized theory
reproduces the QHE:
From the equations of motion,
$\delta S/\delta a_{\mu}=
\delta S/\delta b_{\mu}=0$, 
\begin{align}
&
-
\frac{2\mathsf{k}}{4\pi}
\epsilon^{\mu\nu\lambda} \partial_{\nu} b_{\lambda}
+
\frac{\mathsf{Ch}}{2 \pi} 
\epsilon^{\mu\nu\lambda} \partial_{\nu} a_{\lambda}
=0,
\nonumber \\
&
-\epsilon^{\mu\nu\lambda} \partial_{\nu} 
\left(a_{\lambda}- A^{\mathrm{ex}}_{\lambda}\right)
=0. 
\end{align}
From the definition of the electrical current
\begin{align}
j^{\mu}
:=
\frac{\delta S}{\delta A^{\mathrm{ex}}_{\mu}}
=
\frac{2 \mathsf{k}}{4\pi}
\epsilon^{\mu\nu\lambda}
\partial_{\nu} b_{\lambda}, 
\end{align}
we then conclude the QHE as
\begin{align}
j^{\mu} = 
\frac{\mathsf{Ch}}{2\pi}
\epsilon^{\mu\nu\lambda} \partial_{\nu}A^{\mathrm{ex}}_{\lambda}.
\end{align}

While functional bosonization delivers
the two-component CS theory
(the BF-CS theory),
it is possible to derive the familiar single-component 
CS theory
by integrating over $a_{\mu}$:
If we eliminate $a_{\mu}$ by using the equation of motion,
for the simplest case of 
$\mathsf{k}=\mathsf{Ch}=1$,
we obtain the effective action
\begin{align}
\mathcal{L}
=
-\frac{1}{4\pi} 
\epsilon^{\mu\nu\lambda} b_{\mu} \partial_{\nu} b_{\lambda}
+
\frac{1}{2\pi} 
\epsilon^{\mu\nu\lambda} b_{\mu} \partial_{\nu} 
A^{\mathrm{ex}}_{\lambda}.
\label{cs theory for b}
\end{align}
This is the familiar single-component CS theory
(here for the case of the integer QHE). 
Observe that,
as it should,
the CS coefficient 
in front of
$\epsilon^{\mu\nu\lambda} b_{\mu} \partial_{\nu} b_{\lambda}$
has the opposite sign 
as compared to the
CS theory of the response, 
$(\mathsf{Ch}/4\pi) \epsilon^{\mu\nu\lambda} A^{\mathrm{ex}}_{\mu} \partial_{\nu} A^{\mathrm{ex}}_{\lambda}$.

The transition from 
the BF-Chern Simons-theory
(\ref{bf-cs theory})
to 
the single-component CS theory 
(\ref{cs theory for b})
is quite analogous to 
the derivation of the 
CS description for the (integer) QHE
in terms of the composite boson theory. 
In the composite boson theory, 
one introduces a CS theory which attaches 
a flux (fluxes) to electrons, and the bosonic composite particle
(electron + flux), in terms of a dualized language, 
is expressed by a U(1) gauge field;
These two U(1) gauge fields correspond to $a_{\mu}$ and $b_{\mu}$
in functional bosonization.
In fact,
if we ``dualize back'' 
the BF-Chern Simons-theory
(\ref{bf-cs theory}), 
we obtain the Chern-Siomns-Landau-Ginzburg action,
in terms of
the CS gauge field $a_{\mu}$ and 
and some bosonic field $\Phi$.
Comparing with the composite boson theory,
the bosonic $\Phi$ could be interpreted as a composite boson field.
Thus, in functional bosonization,
while we have not used the flux attachment, 
the resulting effective field theory is quite similar 
to the composite boson theory
of the IQHE.

As is clear from the comparison to the composite boson theory, 
the hydrodynamic theory
\eqref{cs theory for b}
[and hence \eqref{bf-cs theory}]
includes a fermionic excitation (electron)
due to the presence of the Chern-Simons term.
The world-lines of point defects
(point sources for the vortex current that couples to $b_{\mu}$),
when linked, pick up a $\pi$ phase ({\it i.e.}, a fermionic sign)
due to the the Chern-Simons term. 
This should be contrasted with
the hydrodynamic theory for topologically trivial insulators
\eqref{bosonized action D=3 no Ch} which does not have the Chern-Simons term, and thus no fermionic statistics of its excitations.

\subsection{Topological insulator in $D=3+1$}

The bosonized Lagrangian 
Eq.\ (\ref{bosonized action D=4})
describes the $D=3+1$ dimensional (topological) insulator
in symmetry class AIII (DIII). 
As a possible microscopic realization that conserves a U(1) charge,
let us consider the following fermionic action 
\begin{align}
 &
 K_F[\psi^{\dag},\psi, A^{\mathrm{ex}}]
 =\int dt 
 \sum_{r,r'}
\psi^{\dag}(r)
\nonumber \\
&\quad  \times
\left[
  ({i}\partial_{t}+A^{\mathrm{ex}}_{0})\delta_{r,r'}
 -\mathcal{H}(r,r',A^{\mathrm{ex}})\right] 
 \psi(r'),
\end{align}
where $\psi(r)/\psi^{\dag}(r)$ is the fermion annihilation/creation field  operators
at site $r$, and $\mathcal{H}(r,r',A^{\mathrm{ex}})$ represents
a (tight-binding) Hamiltonian of a chiral topological insulator, minimally
coupled to the external electromagnetic U(1) gauge field $A^{\mathrm{ex}}_{\mu}$.
[Here, $\psi(r)$ can possibly be multi-component, but internal indices will be 
suppressed below.]
A single-particle Hamiltonian $\mathcal{H}$ in symmetry class AIII 
is sublattice symmetric in that it 
anticommutes with a unitary matrix $\Gamma$,
$\{\mathcal{H},\Gamma\}=0.$
$\Gamma$ is diagonal in sublattice indices and takes values $\pm 1$
for sublattice A and B, respectively.
This means, in turn, 
under the unitary transformation (particle-hole transformation) 
\begin{align}
& \mathcal{C}\psi(t,r) \mathcal{C}^{-1}=(-1)^{r}\psi^{\dag}(t,r),
\nonumber \\
& (-1)^r =
\left\{
 \begin{array}{ll}
  +1, & r \in \mbox{A sublattice}
\\
-1, & r\in \mbox{B sublattice}
 \end{array}
 \right.  
\end{align}
followed by time-reversal 
\begin{align}
\mathcal{T}\psi(t,r)\mathcal{T}^{-1}=
\psi(-t,r),
\quad
\mathcal{T}{i}\mathcal{T}^{-1}=-{i},
\end{align}
the fermion bilinear
$\int dt\,\sum_{r,r'}\psi^{\dag}(r)\mathcal{H}(r,r',A^{\mathrm{ex}})\psi(r')$
is left unchanged while the sign of $A^{\mathrm{ex}}_{0}$ is flipped, 
{\it i.e.}, $\mathcal{T}\mathcal{C}$
sends
($i=1,2,3$)
\begin{align}
 \mathcal{T}\mathcal{C}:\quad &
 A^{\mathrm{ex}}_0(t,r) \to -A^{\mathrm{ex}}_0 (-t,r),
\nonumber \\
 &
 A^{\mathrm{ex}}_i(t,r) \to +A^{\mathrm{ex}}_{i}(-t,r), 
\end{align}
and
$\vec{E}^{\mathrm{ex}}(t,r)  \to -\vec{E}^{\mathrm{ex}}(-t,r)$, 
$\vec{B}^{\mathrm{ex}}(t,r)  \to + \vec{B}^{\mathrm{ex}}(-t,r)$.

The functional bosonization recipe 
in Sec.\ \ref{sec: Functional bosonization},
when applied to the fermionic action,
delivers the hydrodynamic field theory
(\ref{bosonized action D=4}):
\begin{align}
\mathcal{L}
&=
-
b_{\mu\nu} 
\epsilon^{\mu\nu\lambda\rho} \partial_{\lambda} 
(a_{\rho}-A^{\mathrm{ex}}_{\rho})
\nonumber \\
&\qquad 
+
\frac{\theta}{8\pi^2} 
\epsilon^{\mu\nu\lambda\rho}
\partial_{\mu} a_{\nu}\partial_{\lambda} a_{\rho},
\label{3+1 action}
\end{align}
where we have dropped the 
Maxwell term for simplicity. 
The theta angle $\theta=\nu \pi$ 
where $\nu\in \mathbb{Z}$ 
is the topological invariant (winding number).
This is consistent with chiral symmetry;
because of the transformation law of 
$A^{\mathrm{ex}}$ and similarly $a_{\mu}$ under $\mathcal{T}\mathcal{C}$, 
$\theta$ is mapped to $-\theta$ by $\mathcal{T}\mathcal{C}$.
With the invariance of the the theta term under a shift $\theta\to\theta+2\pi$,
the value of the theta angle allowed by chiral symmetry is
an integral multiple of $\pi$.

From the BF coupling 
$b_{\mu\nu}\epsilon^{\mu\nu\lambda\rho}\partial_{\lambda}A^{\mathrm{ex}}_{\rho}$
or from the bosonization rule
$j^{\mu}\propto \epsilon^{\mu\nu\lambda\rho}\partial_{\nu}b_{\lambda\rho}$, 
one also reads off the transformation law
\begin{align}
 \mathcal{T}\mathcal{C}:\quad &
 b_{0i}(t,r) \to +b_{0i} (-t,r),
 \quad
 b_{i0}(t,r) \to +b_{i0} (-t,r),
\nonumber \\
 & b_{ij}(t,r) \to -b_{ij}(-t,r), 
\end{align}
where $i,j=1,2,3$.

The effective Lagrangian (\ref{3+1 action})
obtained from  
functional bosonization 
was discussed previously
in Refs.\ \onlinecite{Szabo1998,Szabo1999,Diamantini2012},
and (\ref{3+1 action})
should be 
compared with the one proposed in
Ref.\ \onlinecite{ChoMoore2010}:
\begin{align}
\mathcal{L}
&=
\frac{1}{2\pi}
\epsilon^{\mu\nu\lambda\rho}
a_{\mu}\partial_{\nu} b_{\lambda \rho}
+
\frac{1}{2\pi} 
\epsilon^{\mu\nu\lambda\rho}
A^{\mathrm{ex}}_{\mu} \partial_{\nu} b_{\lambda \rho}
\nonumber \\
&
\quad 
+
C 
\epsilon^{\mu\nu\lambda\rho}
\partial_{\mu} a_{\nu} \partial_{\lambda} A^{\mathrm{ex}}_{\rho},
\label{eq: cho-moore}
\end{align}
where $C= \pm 1/(8\pi)$. 
One can check, upon integration 
over $a$ and $b$, we reproduce
the axion response term
\begin{align}
S =
\frac{\pm 1}{8\pi}
\int dt d^3x\,
\epsilon^{\mu\nu\lambda\rho} 
\partial_{\mu} A^{\mathrm{ex}}_{\nu} \partial_{\lambda} A^{\mathrm{ex}}_{\rho}. 
\end{align}
As compared to Eq.\ (\ref{3+1 action})
the effective action (\ref{eq: cho-moore})
does not have the axion term for the $a_{\mu}$ field. 
However,
assuming there is no monopole in $a_{\mu}$ field configurations, 
with the gauge transformation in Eq.\ (\ref{3+1 action})
\begin{align}
b_{\mu\nu}
\to
b_{\mu\nu}
+ 
\frac{\theta}{16\pi^2}
\left(
\partial_{\mu} a_{\nu}
-
\partial_{\nu} a_{\mu}
\right),
\end{align}
one can transform Eq.\ (\ref{3+1 action})
into Eq.\ (\ref{eq: cho-moore}).

As in the case of $D=2+1$, 
one can readily verify that the bosonized effective 
Lagrangian reproduces the physics of 
the topological insulator in $D=3+1$ dimensions:
By eliminating (integrating out) $b_{\mu\nu}$ and $a_{\mu}$  using the equations 
of motion, 
$\delta S/\delta a_{\mu}=
\delta S/\delta b_{\mu\nu}=0$, 
\begin{align}
&
\epsilon^{\mu\nu\lambda\rho}
\partial_{\nu} b_{\lambda\rho}
-
\frac{1}{4\pi^2}
\epsilon^{\mu\nu\lambda\rho}
\partial_{\nu}
\left(
\theta \partial_{\lambda} a_{\rho}
\right)
=0, 
\nonumber \\
&
\epsilon^{\mu\nu\lambda\sigma} \partial_{\lambda} 
(a_{\sigma}- A^{\mathrm{ex}}_{\sigma})
=0, 
\label{3+1 EOM}
\end{align}
one obtains the axion term for the external gauge field
\begin{align}
\frac{1}{8\pi^2} 
\int d^4x\, 
\theta \epsilon^{\mu\nu\rho\sigma}
\partial_{\mu} A^{\mathrm{ex}}_{\nu}\partial_{\rho} A^{\mathrm{ex}}_{\sigma},
\label{3+1 axion free}
\end{align}
as expected for the response of the topological insulator
in symmetry classes AIII and DIII in $D=4$. 
Also, from the bosonization rule
$
j^{\mu} 
\equiv
\epsilon^{\mu\nu\lambda\rho}
\partial_{\nu} b_{\lambda\rho}
$, 
the electrical current is given by
\begin{align}
j^{\mu} 
&=
\epsilon^{\mu\nu\lambda\rho}
\partial_{\nu} b_{\lambda\rho}
=
\frac{1}{4\pi^2}
\epsilon^{\mu\nu\lambda\rho}
\partial_{\nu}
\left(
\theta \partial_{\lambda} A^{\mathrm{ex}}_{\rho}
\right). 
\label{current in ti}
\end{align}
Taking $\theta$ to be a step function which jumps
from $\pi \to 0$ at the system's boundary,
this equation
tells us that surface Hall conductance is
$\sigma_{xy}=
1/4\pi= e^2/2h$ where we have reinserted units.
Alternatively, when $\theta=\pi=\mbox{const.}$,
but $A^{\mathrm{ex}}_{\rho}$ has a monopole configuration, 
this equation tells us 
the monopole acquires a charge which is a manifestation of the
Witten effect.
\cite{Witten1979,Qi2009,Rosenberg2010}

To discuss the effect of monopoles, 
it is convenient to introduce
{\it a monopole gauge field} 
$f^{\mathrm{M}}_{\mu\nu}[A^{\mathrm{ex}}]$
and
$f^{\mathrm{M}}_{\mu\nu}[a]$
for 
$A^{\mathrm{ex}}_{\mu}$
and
$a_{\mu}$, 
respectively.
\cite{Kleinert,Quevedo1997,Diamantini2012}
In the presence of a monopole
in a U(1) gauge field
$A_{\mu}$ (it can be either $A^{\mathrm{ex}}_{\mu}$
or $a_{\mu}$),
the monopole gauge field
$f^{\mathrm{M}}_{\mu\nu}[A]$
is a field which is related to 
the monopole current 
$j^{\mathrm{M},\mu}$
as 
\begin{align}
\frac{1}{2} \epsilon^{\mu\nu\lambda\rho}
\partial^{\ }_{\nu}f^{\mathrm{M}}_{\lambda\rho}[A]
=
j^{\mathrm{M},\mu}. 
\label{monopole gauge field and current}
\end{align}
More specifically, it is given by
\begin{align}
f^{\mathrm{M}}_{\mu\nu}[A](x)
&=
\frac{q_m}{2}
\epsilon_{\mu\nu\lambda\rho}
\delta^{\lambda\rho}(x;S)
=
q_m \tilde{\delta}_{\mu\nu}(x;S), 
\end{align}
where
$q_m$
is the strength of the monopole, 
and 
$\delta_{\lambda\rho}(x;S)$
is singular on the world-surface $S$, 
\begin{align}
\delta^{\mu\nu}(x;S)
&:=
\int d \sigma d\tau
\left[
\frac{ \partial X^{\mu}(\sigma,\tau)}{\partial \sigma}
\frac{ \partial X^{\nu}(\sigma,\tau)}{\partial \tau}
-
\left(
\mu \leftrightarrow \nu
\right)
\right]
\nonumber \\
&\quad
\times 
\delta^{(4)}[x - X(\sigma,\tau)].
\end{align}
The surface $S$ is the world-sheet of the Dirac string,
parameterized by its world-sheet coordinates $X^{\mu}(\sigma,\tau)$, 
and bounded by a world-line $L$
of the monopole, $\partial S=L$. 
One verifies 
\begin{align}
\frac{1}{2}
\epsilon^{\mu\nu\lambda\rho}
\partial_{\nu}\tilde{\delta}_{\lambda\rho}(x;S)
=
\delta^{\mu} (x;L),
\end{align}
where
\begin{align}
\delta^{\mu}(x;L)
=
\int d \sigma
\frac{ \partial X^{\mu}(\sigma,\tau)}{\partial \sigma}
\delta^{(4)}[x - X(\sigma)].
\end{align}
One then concludes 
$f^{\mathrm{M}}_{\mu\nu}[A^{\mathrm{ex}}]$
is related to the monopole current as
Eq.\ (\ref{monopole gauge field and current}).
It represents the
magnetic field inside the infinitely thin
solenoid that eminates from the monopole. 

The physically observable field strength,
$f^{\mathrm{obs}}_{\mu\nu}[A]$, 
is the difference of the 
field strength
for the {\it integrable} vector potential $A_{\mu}$
and the monopole gauge field,
$f^{\mathrm{obs}}_{\mu\nu}[A]
= f_{\mu\nu}[A]
-f^{\mathrm{M}}_{\mu\nu}[A]$.
The curl of the observable field strength 
$f^{\mathrm{obs}}_{\mu\nu}[A]$
is non-zero in the presence of
a magnetic monopole,  
$(1/2) \epsilon^{\mu\nu\lambda\rho}
\partial_{\nu}f^{\mathrm{obs}}_{\lambda\rho}[A] = 
-j^{\mathrm{M},\mu}
$,
as expected, 
where we noted the integrability of $A_{\mu}$,
$\left(
\partial_{\mu}\partial_{\nu}
-
\partial_{\nu}\partial_{\mu}
\right)
A_{\lambda}=0$.
While both $A_{\mu}$
and $f^{\mathrm{M}}_{\mu\nu}[A]$ depend on 
the choice of the world surface $S$
({\it i.e.}, for a given Dirac string $L$, there are many surfaces satisfying
$\partial S=L$), 
$f^{\mathrm{obs}}_{\mu\nu}[A]$
does not. 
In other words,
$f^{\mathrm{obs}}_{\mu\nu}[A]$ is 
invariant under
the {\it monopole gauge transformation}, 
\begin{align}
f^{\mathrm{M}}_{\mu\nu}[A]
&\to
f^{\mathrm{M}}_{\mu\nu}[A]
+
\partial_{\mu}\eta_{\nu}
-
\partial_{\nu}\eta_{\mu},
\nonumber \\
A_{\mu}
&\to
A_{\mu}
+
\eta_{\mu}. 
\label{monopole gauge trsft}
\end{align}

The partition function, 
invariant under both
the ordinary U(1) gauge transformation
$A_{\mu}\to A_{\mu}+\partial_{\mu}\lambda$
and the monopole gauge transformation
(\ref{monopole gauge trsft}), 
can be constructed from the Lagrangian
\begin{align}
\mathcal{L}
&=
-
\frac{1}{2}b_{\mu\nu} 
\epsilon^{\mu\nu\lambda\rho}
\left(
f_{\lambda\rho}[a]-
f_{\lambda\rho}[A^{\mathrm{ex}}]
\right)
\nonumber \\
&\quad 
+
\frac{\theta}{32\pi^2} 
\epsilon^{\mu\nu\lambda\rho}
(f_{\mu\nu}[a] -f^{\mathrm{M}}_{\mu\nu}[a])
(f_{\lambda\rho}[a]-f^{\mathrm{M}}_{\lambda\rho}[a]).
\label{3+1 action with monopole}
\end{align}
As opposed to the second term (the axion term),
the first term (the BF term) in Eq.\ (\ref{3+1 action with monopole})
is not manifestly invariant under the monopole gauge transformation. 
However, 
given the bosonization rule
$\epsilon^{\mu\nu\lambda\rho\sigma}\partial_{\nu} b_{\lambda\rho\sigma}\propto j^{\mu}$, 
Dirac's quantization condition
of the electric and magnetic charges
tells us that the BF term 
is merely shifted by an integer multiple of $2\pi$
under the monopole gauge transformation, 
and hence the 
partition function is not affected.

Let us now focus on the case 
where $\theta$ is constant throughout the bulk. 
We can use the equations of motion
$\delta S/\delta a_{\mu}=
\delta S/\delta b_{\mu\nu}=
0$ 
derived from the effective action
(\ref{3+1 action with monopole}) to find that
the electrical current is given in terms of 
the monopole current as 
\begin{align}
j^{\mu}
&=
\epsilon^{\mu\nu\lambda\rho}
\partial_{\nu} b_{\lambda\rho}
=
-\frac{\theta}{8\pi^2}
\epsilon^{\mu\nu\lambda\rho}
\partial_{\nu}
f^{\mathrm{M}}_{\lambda\rho}[A^{\mathrm{ex}}]
\nonumber \\
&=
-\frac{\theta}{4\pi^2}
j^{\mathrm{M},\mu}.
\label{flux attach 4D}
\end{align}
Given that
in $D=2+1$ the gauge field $a_{\mu}$ plays the role 
of attaching a flux (fluxes) to electrons,
it is tempting, and perhaps instructive, to view 
the gauge field $a_{\mu}$ in $D=3+1$ 
as a proper generalization of the CS ``statistical gauge field.''
From the equation of motion
$
\epsilon^{\mu\nu\lambda\rho}
\partial_{\nu} b_{\lambda\rho}
=
-(\theta/4\pi^2) j^{\mathrm{M},\mu}
$,
the gauge field $a_{\mu}$ is attaching
a monopole to the electron
(and as in $D=2+1$,
the temporal component $a_{0}$ enforces a constraint).
If we further choose, in the presence of a vortex line,
the worldsheet of the Dirac string
to be identical to 
the vortex worldsheet,
the boundary of the vortex worldsheet
(the end of the vortex)
is the worldline of a monopole 
({\it i.e.}, a dyon since $\theta\neq 0$).
In this case,
Eq.\ (\ref{flux attach 4D})
gives 
$   b_{\mu\nu} =
    -(\theta/8\pi^2)f^{\mathrm{M}}_{\mu\nu}[A^{\mathrm{ex}}]
$ up to a choice of gauge.

\section{Dimensional reduction}
\label{sec: Dimensional reduction}

We have derived
the effective BF topological field theories
for the primary series of the topological insulators in the periodic table.
In microscopic fermionic theories,
$\mathbb{Z}_2$ topological insulators
(such as the time-reversal symmetric $\mathbb{Z}_2$
topological insulator in $D=3+1$ dimensions, and
the quantum spin Hall effect in $D=2+1$ dimensions)
can be derived
as a ``descendant''of the primary series by
the Kaluza-Klein dimensional reduction. 
\cite{Qi_Taylor_Zhang2008}
We now discuss
the Kaluza-Klein dimensional reduction
of the BF type effective action we derived.
This is expected to deliver
an effective bosonic field theory description for
descendant $\mathbb{Z}_2$ topological insulators. 
Similar
dimensional reduction was discussed
in Ref.\ \onlinecite{Qi_Taylor_Zhang2008}
for the topological \emph{response} theories.

Let us start from 
the $D=4+1$ dimensional BF-CS theory
(the ``parent'' theory): 
\begin{align}
S
&=
-\frac{1}{2\pi}\int d^5x\, 
b_{\mu\nu\lambda} 
\epsilon^{\mu\nu\lambda\rho\sigma} \partial_{\rho} 
(a_{\sigma}-A^{\mathrm{ex}}_{\sigma})
\nonumber \\
&\quad
+
\frac{\mathsf{Ch}_2}{24\pi^2}
\int d^5 x\, 
\epsilon^{\mu\nu\lambda \rho\sigma}
a_{\mu} 
\partial_{\nu}
a_{\lambda} 
\partial_{\rho}
a_{\sigma}
+
\cdots,  
\label{parent, 5D}
\end{align}
where $\mu,\nu,\lambda,\ldots=0,1, 2, 3, 4$. 
The topological insulator in $D=4+1$ dimensions 
is time-reversal symmetric (symmetry class AII)
and so is the effective hydrodynamic theory;
the action (\ref{parent, 5D}) is invariant under 
time-reversal defined by 
($i,j,k=1,\ldots,4$)
\begin{align}
 \mathcal{T}:\quad &
 a_{0}(t,r)\to +a_{0}(-t,r),
 \quad
 a_{i}(t,r)\to -a_{i}(-t,r),
 \nonumber \\
 &
 b_{0ij}(t,r) \to -b_{0ij} (-t,r),
 \quad
 b_{ijk}(t,r) \to +b_{ijk}(-t,r). 
\end{align}

We now expand the fields as
\begin{align}
\Phi(x_{\mu})
=
\sum^{+\infty}_{n_w=-\infty}
e^{ {i} 2 \pi n_w w/L_w } 
\Phi(x_{i},n_w)
\label{fourier}
\end{align}
where $\Phi$ represents a field,
$L_w$ is the circumference of $x^4$ direction,
and $x_i=(x_0,x_1,x_2,x_3)$ and $x_4=w$.  
By ``shrinking'' the $x^4$ direction by taking $L_w\to 0$,
the ``parent''  $D=4+1$ dimensional theory is reduced to
a descendant theory in $D=3+1$ dimensions.
In doing so, we view each Fourier mode
$\Phi(x_{i},n_w)$ as a field in $D=3+1$ dimensions
(Kaluza-Klein modes). 
The modes with $n_w \neq 0$ have a gap which grows as we take
$L_w\to 0$ and hence at low energies only the Fourier modes with
$n_w=0$ are important.
[This can be seen in the presence of proper kinetic terms
({\it e.g.},  the Maxwell term)
for $a_{\mu}$ and $b_{\mu\nu\lambda}$, which are not shown explicitly above]. 
We define 
\begin{align}
a_{w}(x_i, n_w=0)
&=: \phi(x_i)/L_w,
\nonumber \\
b_{wij}(x_i, n_w=0)
&=: 
\frac{1}{3} u_{ij}(x_i)/L_w,
\nonumber \\
A^{\mathrm{ex}}_{w}(x_i, n_w=0)
&=: \theta^{\mathrm{ex}}(x_i)/L_w,  
\label{dim red 1}
\end{align}
($i,j=0,\ldots,4$). 
The resulting $D=3+1$-dimensional Lagrangian
inherits the field content of the parent theory:
1-form gauge fields $a_{i}$ and $A^{\mathrm{ex}}_{i}$,
a 3-form gauge field $b_{ijk}$,
where the space-time index in $D=4$ dimensions
$i,j,k$ runs from $0$ to $3$. 
In addition, as introduced in Eq.\ (\ref{dim red 1}),
it also contains two scalar fields,
$\phi$ and $\theta^{\mathrm{ex}}$, and
one 2-form gauge field $u_{ij}$. 
The scalar field
$\theta^{\mathrm{ex}}$ is the background axion field,
which satisfies $\theta^{\mathrm{ex}}=\pi (0)$ inside (outside) of the
$D=3+1$ dimensional topological insulator.
Putting together, the effective topological field theory
for the $D=3+1$ dimensional topological insulator 
is given by the Lagrangian, 
\begin{align}
\mathcal{L}
&=
-
\frac{\epsilon^{\mu \nu \lambda \rho}}{2\pi}
(\phi-\theta^{\mathrm{ex}})
\partial_{\mu} b_{\nu \lambda \rho}
-
\frac{\epsilon^{\mu\nu\lambda\rho}}{2\pi}
(a_\mu-A^{\mathrm{ex}}_\mu) \partial_\nu u_{\lambda\rho}
\nonumber \\
&\quad
+
\frac{\mathsf{Ch}_2}{8\pi^2}
\epsilon^{\mu \nu \lambda \rho}
\phi
\partial_{\mu}
a_{\nu} 
\partial_{\lambda}
a_{\rho}
+
\cdots,  
\end{align}
where the space-time index $\mu,\nu,\lambda$ now runs from $0$ to $4$. 
The last two terms resemble
the effective field theory of the $D=3+1$ dimensional topological
insulator in the primary series, 
{\it i.e.,}, the BF theory with the axion term,
Eq.\ (\ref{bosonized action D=4}),
whereas the first term is absent in
Eq.\ (\ref{bosonized action D=4}).
The $\mathbb{Z}_2$ nature of the system lies in the restriction 
on $\mathsf{Ch}_2 \theta^{\mathrm{ex}}$ 
which is quantized by time-reversal symmetry to be fixed 
at $\mathsf{Ch}_2\theta^{\mathrm{ex}}=2n\pi$ or $(2n+1)\pi$ where 
$n$ is an integer. 
The $\mathbb{Z}_2$ nature is manifest in the fact that the bulk of the material only uniquely determines whether  $\mathsf{Ch}_2 \theta^{\mathrm{ex}}$ is an even or odd multiple of $\pi$ not what the value of $n$ is.\cite{Qi_Taylor_Zhang2008}
If we calculate the electrical current response we find
\begin{align}
\frac{\delta{S}}{\delta A^{\mathrm{ex}}_{\mu}}
&=
j^{\mu}
=
\frac{1}{2\pi}\epsilon^{\mu\nu\lambda\rho} \partial_{\nu} u_{\lambda\rho}, 
\nonumber \\
\frac{\delta{S}}{\delta a_{\mu}}
&=
- \frac{1}{2\pi}\epsilon^{\mu\nu\lambda\rho} \partial_{\nu} u_{\lambda\rho}
+
\frac{2\mathsf{Ch}_2}{8\pi^2}
\epsilon^{\mu \nu \lambda\rho }
\partial_{\nu}
\left(
\phi 
\partial_{\lambda} a_\rho
\right)=0,
\nonumber\\
\frac{\delta{S}}{\delta b_{\nu\lambda\rho}}
&=
\frac{1}{2\pi}\epsilon^{\mu\nu\lambda\rho} \partial_{\mu}(\phi- \theta^{\mathrm{ex}})=0,\nonumber\\
\frac{\delta{S}}{\delta u_{\lambda\rho}}
&=
\frac{1}{2\pi}\epsilon^{\mu\nu\lambda\rho} \partial_{\nu}(a_{\mu}-A^{\mathrm{ex}}_{\mu})=0,
\end{align}
which gives rise to 
\begin{align}
j^{\mu} 
=
\frac{\mathsf{Ch}_2}{4\pi^2}
\epsilon^{\mu \nu \lambda\rho }
\partial_{\nu}\theta^{\mathrm{ex}}
\partial_{\lambda} A^{\mathrm{ex}}_{\rho}
\end{align}
where we have ignored possible monopole contributions to the current. This response implies that if $\mathsf{Ch}_2\theta^{\mathrm{ex}}=(2n+1)\pi$ inside the material and $2n\pi$ outside then there is a half-integer quantum Hall effect on the surface.

We can continue the reduction down to the second descendant in $D=2+1$
to obtain 
a hydrodynamic effective field theory for
the quantum spin Hall effect;
we separate the $D=4$-dimensional coordinates
$(x_0, x_1, x_2,x_3)$
into the $D=3$-dimensional ones
$(x_0, x_1, x_2)$
and 
$x_3 \equiv z$.
The $z$-direction is compactified on a circle
with radius $L_z$ and then we take $L_z\to 0$.
The fields can be Fourier decomposed in the
$z$-direction, as in Eq.\ (\ref{fourier}),
and only the Fourier modes with $n_z=0$
(the momentum quantum number in the $z$-direction)
are kept.
This second step  of dimensional reduction
introduces,
in addition to
two new scalar fields $\psi, \chi^{\mathrm{ex}}$,
a vector field $v_i$,
and
a second 2-form gauge field $g_{ij}$
($i,j=0,1,2,3$). 
They are defined, from the $D=3+1$ dimensional fields, 
as 
\begin{align}
a_{z}(x_i, n_z)
&=: \psi(x_i)/L_z,
\nonumber \\
b_{zij}(x_i, n_z)
&=: 
\frac{1}{3} g_{ij}(x_i)/L_z,
\nonumber \\
u_{zi}(x_i, n_z)
&=:
\frac{1}{2} 
v_{i}(x_i)/L_z,
\nonumber \\
A^{\mathrm{ex}}_{z}(x_i, n_z)
&=: \chi^{\mathrm{ex}}(x_i)/L_z. 
\end{align}
The resulting $D=2+1$ dimensional Lagrangian is 
\begin{align}
\mathcal{L}
&=
-\frac{\epsilon^{\mu \nu \lambda}}{2\pi}
(\phi-\theta^{\mathrm{ex}})
\partial_{\mu} g_{\nu \lambda}
-
\frac{\epsilon^{\mu \nu \lambda}}{2\pi}
(\psi-\chi^{\mathrm{ex}}) \partial_{\mu} u_{\nu \lambda}
\nonumber \\
&
\quad
-
\frac{\epsilon^{\mu \nu \lambda}}{2\pi}
(a_\mu-A^{\mathrm{ex}}_\mu) \partial_\nu v_{\lambda}
+
\frac{\mathsf{Ch}_2}{4\pi^2}
\epsilon^{\mu \nu \lambda}
\phi 
\partial_\mu \psi
\partial_{\nu}
a_{\lambda}, 
\label{eff FT for QSHE}
\end{align}
where the space-time index $\mu,\nu,\lambda$ now runs from $0$ to $2$. 
We recognize the third term
$\epsilon^{\mu \nu \lambda}
(a_\mu-A^{\mathrm{ex}}_\mu) \partial_\nu v_{\lambda}$
as the BF coupling in $D=2+1$ dimensions with $b_\mu$ replaced by $v_\mu.$
The presence of such term
is largely expected based upon
the phenomenology of the non-chiral (helical) edge modes of the quantum spin Hall effect.
However, there are some additional terms which have not been previously discussed in dynamical gauge theories: the coupling to the real scalar fields $\phi$ and $\psi$ which could be combined into a single complex scalar field. We note that this theory, unlike other hydrodynamic theories of the quantum spin Hall effect
\cite{Bernevig2006,Levin2009,Levin2012,Neupert2011,Santos2011,Lu2012}
only has a single $\mathrm{U}(1)$ gauge invariance 
(unlike the more conventional $\mathrm{U}(1)\times \mathrm{U}(1)$ for charge and spin).
This is a natural result since in real materials spin is not conserved generically and only the local charge $\mathrm{U}(1)$ invariance is preserved.

It is a simple exercise to read off,
from the above effective field theories,
a response of the system to the external field.
For example, for $D=2+1$ dimensions,
the definition of the electrical current
$j^{\mu}:=\delta{S}/\delta A^{\mathrm{ex}}_{\mu}$
together with the equation of motions
derived from the effective action (\ref{eff FT for QSHE}), 
\begin{align}
\frac{\delta{S}}{\delta A^{\mathrm{ex}}_{\mu}}
&=
j^{\mu}
=
\frac{\epsilon^{\mu\nu\lambda}}{2\pi} \partial_{\nu} v_{\lambda}, 
\nonumber \\
\frac{\delta{S}}{\delta a_{\mu}}
&=
- \frac{\epsilon^{\mu\nu\lambda}}{2\pi} \partial_{\nu} v_{\lambda}
+
\frac{\mathsf{Ch}_2}{4\pi^2}
\epsilon^{\mu \nu \lambda }
\partial_{\nu}
\left(
\phi 
\partial_{\lambda} \psi
\right)=0,
\end{align}
gives rise to 
\begin{align}
j^{\mu} 
=
\frac{\mathsf{Ch}_2}{4\pi^2}
\epsilon^{\mu \nu \lambda }
\partial_{\nu}
\left(
\phi 
\partial_{\lambda} \psi
\right).
\end{align}
 If we assume that 
$\epsilon^{\mu\nu}\partial_{\mu}\partial_{\nu}\psi=0$
 then we can write this in terms of the external scalar fields as
\begin{align}
j^{\mu} 
=
\frac{\mathsf{Ch}_2}{4\pi^2}
\epsilon^{\mu \nu \lambda }
\partial_{\nu}\theta^{\mathrm{ex}} 
\partial_{\lambda} \chi^{\mathrm{ex}}.
\end{align}
A similar expression for the electrical current
was derived in Ref.\ \onlinecite{Qi_Taylor_Zhang2008}
in terms of the effective topological response theory. If we think of the field $\phi\partial_\lambda \psi$ as a velocity field $V_\lambda$ then this response implies that there is charge bound to the vorticity of the velocity
field $j^0 = (\mathsf{Ch}_2/ 4\pi^2)
\nabla\times {\bf{V}},$ similar to the quantum Hall effect where charge is bound to the magnetic flux.
A simple example is the case when
$\theta^{\mathrm{ex}}(x,y)=\pi \Theta(y)$ and 
$\chi^{\mathrm{ex}}(x,y)=2\pi[\Theta(x)-1/2]$ where $\theta^{\mathrm{ex}}$ represents a jump in the the axion angle at an edge and $\chi^{\mathrm{ex}}$ represents a magnetic domain wall on the edge. The charge confined to the edge magnetic domain wall is  $Q=
( e\mathsf{Ch}_2/4\pi^2) \int dx dy\, 2\pi^2 \delta(x)\delta(y)
= e\mathsf{Ch}_2/2.$
The distinction between integer and half-integer charge, \emph{i.e.} $\mathsf{Ch}_2$ even or odd, gives an electromagnetic characteristic to determine the
$\mathbb{Z}_2$ topological nature of the 2d time-reversal invariant topological insulator.
\cite{Qi_Taylor_Zhang2008} Another consequence of the action is the fermionic mutual statistics between quasi-particles that couple as $j^{\mu}a_\mu$ and $K^{\mu} v_{\mu}.$ When $\mathsf{Ch}_2$ vanishes the currents $j^{\mu}, K^{\mu}$ have mutual fermionic statistics but when $\mathsf{Ch}_2\neq 0$ the current $j^{\mu}$ is shifted such that $\tilde{j}^{\mu}$ and $K^{\mu}$ carry mutual fermionic statistics where
\begin{equation}
\tilde{j}^{\mu}=j^{\mu}-\frac{\mathsf{Ch}_2}{4\pi^2}\epsilon^{\mu\nu\lambda}\partial_\nu(\phi\partial_{\lambda}\psi).
\end{equation}

\section{Fractional states}
\label{sec: Fractional states}

Our discussions so far concern a bosonized description of
non-interacting topological band insulators, or
weakly interacting topological insulators
that are adiabatically connected to a topological band insulator.
In these systems, functional bosonization gives rise to the BF topological
field theory with the unit level $\mathsf{k}=1$.
In this section,
we will discuss the possibility of topological states
that can be induced by strong interactions
and characterized by, {\it e.g.}, non-zero ground state degeneracy.
While the existence of strongly interacting
topological insulators with topological order
is not well established microscopically for $D>3$,
we can use the hydrodynamic bosonization formalism of the preceding
sections to explore possible mechanisms that lead
to a topological order.
One such route is the
fractionalization of electrons due to strong correlations. 
We will implement this through the parton construction a la
Blok and Wen, 
\cite{Blok1990a, Blok1990b}
to find effective topological field theories.

\subsection{Parton Construction for the Chern Insulator in $D=2+1$}
We start with a
construction of time-reversal breaking fractional states
in $D=2+1$ dimensions
by combining functional bosonization and the parton construction.
When applied to elections in the (lowest) Landau level,
this approach is equivalent to 
the composite particle (composite boson and composite fermion)
theories.
\cite{
ZhangHanssonKivelson1989,
LeeZhang1991, 
Zhang1992, 
ZeeBook,Wen1990,Wen1992,WenBook,Wen1995,
Jain89a,Jain89b,LopezFradkin1991,FradkinBook}
The functional bosonization is also readily applicable to
fractional quantum Hall states formed on a lattice,
{\it i.e.,} 
``fractional Chern insulators''.
See Refs.\
\onlinecite{Neupert11,Sheng11,Wang11,Regnault11,
LuRan2012,McGreevy2012,
Venderbos2012, 
Wu2012,
Qi2012,
Parameswaran2012,
Bernevig2012,
Wu2012-06,
Wu2012-10} 
for recent studies on
the fractional Chern insulators.

We focus on
the effective field theory
for the hierarchy states at 
$\nu= \mathsf{m}/(\mathsf{m}\mathsf{p}+1)$
in terms of the parton construction. 
Following Refs.\ \onlinecite{Blok1990a, Blok1990b}
we first split the electron
into $\mathsf{p}+1$ partons. 
Here, $\mathsf{p}$ is an even integer, 
and we require partons to obey Fermi statistics. 
Since the electrons do not split in reality, 
we impose a constraint 
\begin{align}
j^{(i)}_{\mu} 
=
j^{(j)}_{\mu},
\quad
i,j = 1,\ldots, \mathsf{p}+1,   
\end{align}
on the parton densities
$j^{(i)}_{\mu}$. 
The $i$-th parton carries electric charge $e_i$,
and
$e= \sum_i e_i =1$. 
We assume the all partons are in 
independent integer quantum Hall states, but we treat
the $i=1,\ldots, \mathsf{p}$-th flavors
and
the $i=\mathsf{p}+1$-st flavor 
differently. 
For $i=1,\ldots, \mathsf{p}$, 
we assume the filling faction is $\nu^{(i)}= 1$
whereas for $i=\mathsf{p}+1$, 
$\nu^{(i)}=\mathsf{m}$.
The total filling fraction is
\begin{align}
\nu =
\frac{1}{
\mathsf{p} + 1/\mathsf{m}
}
=
\frac{\mathsf{m}}{
\mathsf{m}\mathsf{p} + 1
}.
\end{align}

Each parton can be bosonized by functional bosonization. 
For
$i=1,\ldots, \mathsf{p}$-th partons,
they can be described by the following
$D=2+1$ BF theories with the CS term, 
\begin{align}
\mathcal{L}^{(i)}
&=
\frac{-1}{2\pi}
\epsilon b^{(i)}\partial a^{(i)}
+
\frac{1}{4\pi} 
\epsilon a^{(i)} \partial a^{(i)}
-e_i j^{(i)} \cdot A^{\mathrm{ex}},  
\end{align}
where the repeated indices $i$ are not summed,
and we have introduced a short hand notations
$A^{\mathrm{ex}}\cdot j
\equiv  A^{\mathrm{ex}}_{\mu} j^{\mu}
$
and
$\epsilon a\partial a
\equiv
\epsilon^{\mu\nu\lambda} a_{\mu} \partial_{\nu} a_{\lambda}$,
{\it etc.} 
On the other hand, for the $i=\mathsf{p}+1$-st parton, 
since we have $\mathsf{m}$ filled Landau levels, 
we introduce
$\mathsf{m}$ separate gauge fields,
$b^{I=1,\ldots,\mathsf{m}}_{\mu}$
and
$a^{I=1,\ldots,\mathsf{m}}_{\mu}$, 
each representing the condensate in the $I$-th Landau level. 
The Lagrangian for
the $i=\mathsf{p}+1$-st parton is 
\begin{align}
\mathcal{L}^{(\mathsf{p}+1)}
&=
\sum_{I=1}^{\mathsf{m}}
\left[
\frac{-1}{2\pi} 
\epsilon b^{I}\partial a^{I}
+
\frac{1}{4\pi} 
\epsilon a^{I} \partial a^{I}
-e_{\mathsf{p}+1} 
j^{I} \cdot A^{\mathrm{ex}}
\right].  
\end{align}
The total Lagrangian is given by
$\mathcal{L}=\sum_{i=1}^{\mathsf{p}}\mathcal{L}^{(i)}
+ \mathcal{L}^{(\mathsf{p}+1)}$.

With the constraints imposed on the parton densities, 
\begin{align}
  \epsilon^{\mu\nu\lambda}
  \partial_{\nu}
b^{(i)}_{\lambda} 
=
  \epsilon^{\mu\nu\lambda}
  \partial_{\nu}
b^{(\mathsf{p}+1)}_{\lambda}
=
\sum_I
  \epsilon^{\mu\nu\lambda}
  \partial_{\nu}
b^{I}_{\lambda},
\end{align}
($i=1,\ldots,\mathsf{p}$), 
the total Lagrangian (after solving the constraints)
is given by
\begin{align}
\mathcal{L}
&=
\frac{-1}{2\pi} 
\sum_{I}  
\epsilon b^{I} \partial a^{I}
+
\frac{-1}{2\pi} 
\sum_{I,i}  
\epsilon b^{I} \partial a^{{(i)}}
\nonumber \\
&\quad
+
\frac{1}{4\pi}
\sum_{I}  
\epsilon a^{I} \partial a^I
+
\frac{1}{4\pi}
\sum_{i}  
\epsilon a^{(i)} \partial a^{(i)}
\nonumber \\
&\quad 
	-\frac{e}{2\pi} \sum_I \epsilon \partial b^I A^{\mathrm{ex}}.
\label{parton lagrangian 2}
\end{align}
This can be written more compactly as
\begin{align}
\mathcal{L} &=
	\sum_{i,j}\frac{\tilde{K}_{ij}}{4\pi}
	\epsilon \alpha^{i} \partial \alpha^j
		-\sum_{i} \frac{e}{2\pi} q_{i} \epsilon \partial \alpha^i A^{\mathrm{ex}}, 
	\label{large K matrix theory}
\end{align}
where
$\alpha=(b^I,a^I,a^{(i)})$,
the charge vector $q_i$ is given as
$q_i=1$ for the first $\mathsf{m}$ entries
whereas
$q_i=0$ otherwise,
and
the $K$-matrix is
\begin{align}
\tilde{K} =
\left(
\begin{array}{ccc}
0 & -I_{\mathsf{m}} &  -J_{\mathsf{m},\mathsf{p}}
\\
-I_{\mathsf{m}} & I_{\mathsf{m}} & 0 
\\ 
-J_{\mathsf{p},\mathsf{m}} & 0 & I_{\mathsf{p}}
\end{array}
\right)
\end{align}
where  $I_{\mathsf{m}}$ is an $\mathsf{m}\times \mathsf{m}$ identity matrix and $J_{\mathsf{m},\mathsf{p}}$ is the $\mathsf{m}\times \mathsf{p}$ matrix
with all matrix elements = 1.

The topological order encoded in
the $\tilde{K}$-matrix CS theory
\eqref{large K matrix theory}
can equivalently be described by
a multi-component CS theory with fewer components. 
For example, when $\mathsf{m}=1$ and $\mathsf{p}=2$,
the filling fraction is $\nu=1/3$ and
\begin{align}
\tilde{K} =
\left(
\begin{array}{cccc}
0 & -1 &  -1 & -1
\\
-1 & 1 & 0 & 0
\\ 
-1 & 0 & 1 & 0
\\
-1 & 0 & 0 & 1
\end{array}
\right). 
\end{align}
The ground state degeneracy can be read off from
$
|\mathrm{Det}\, \tilde{K}|
=
3  
$,
as expected for the Laughlin state at $\nu=1/3$. 
We can integrate out the $a^{(i)}$ one by one
using the equations of motion to arrive at the more familiar form of the $K$-matrix: 
With the equations of motion
$\delta S/\delta a^{(i)}_{\mu}=
-(2\pi)^{-1}
\epsilon^{\mu \nu\lambda}
\partial_{\nu} b_{\lambda}
+
(2\pi)^{-1}
\epsilon^{\mu \nu\lambda}
\partial_{\nu} a^{(i)}_{\lambda}
=0,
$
then, the Lagrangian 
\begin{align}
\mathcal{L}
&=
\frac{-1}{2\pi} 
\epsilon b \partial a
+
\frac{-1}{2\pi} 
\sum^{\mathsf{p}}_{i=1}  
\epsilon b \partial a^{{(i)}}
\nonumber \\
&\quad
+
\frac{1}{4\pi}
\epsilon a \partial a
+
\frac{1}{4\pi}
\sum^{\mathsf{p}}_{i=1}
\epsilon a^{(i)} \partial a^{(i)}
\nonumber \\
&\quad
-\frac{e}{2\pi} \epsilon \partial b A^{\mathrm{ex}}
\label{parton lagrangian reduced 1}
\end{align}
can be reduced to
\begin{align}
  \mathcal{L}
  &\to
\frac{-1}{2\pi} 
\epsilon b \partial a
+
\frac{1}{4\pi}
\epsilon a \partial a
\nonumber \\
&\quad 
+
\frac{-\mathsf{p}}{2\pi} 
\epsilon b \partial a^{{(\mathsf{p})}}
+
\frac{\mathsf{p}}{4\pi}
\epsilon a^{(\mathsf{p})} \partial a^{(\mathsf{p})}. 
\nonumber \\
&\quad
-\frac{e}{2\pi} \epsilon \partial b A^{\mathrm{ex}}. 
\label{parton lagrangian reduced 2}
\end{align}
This is identical to the effective action obtained 
from the flux attachment composite particle approach,
where $b_{\mu}$ plays the role of the vortex gauge field
whereas $a_{\mu}$ is the statistical CS gauge field that
transmutes the statistics of the electrons.

\subsection{Parton Construction for Topological Insulators in $D=3+1$} 
We can formally repeat the parton construction
in $D=3+1$. 
As before, 
we postulate that 
electrons are fractionalized,
consist of 
$\mathsf{p}+1$ partons,
and
each parton is in its topological insulator phase.
For each parton, we can apply functional bosonization
to derive its hydrodynamic theory. 
Solving the constraints among parton densities,
we will arrive at multi-component BF theories
with a topological term (\emph{e.g.} axion or $\theta$-term in $D=3+1$).
In this section, we assume partons are in
a $D=3+1$ dimensional topological insulator phase
in symmetry class AIII or DIII characterized by
an integer topological invariant. 
See Refs.\ \onlinecite{Maciejko10,Maciejko11,Swingle10}
for previous studies of
time-reversal symmetric 
fractional topological insulators
in $D=3+1$ in terms of the parton construction.

We thus write down 
the following Lagrangian 
$\mathcal{L}= \sum_{i=1}^{\mathsf{p}} \mathcal{L}^{(i)} + \mathcal{L}^{(\mathsf{p}+1)}$
for partons
[see Eq.\ \eqref{3+1 action}],  
\begin{align}
\mathcal{L}^{(i)}
&=
-
\epsilon^{\mu\nu\lambda\rho}
b^{(i)}_{\mu\nu} \partial_{\lambda}a^{(i)}_{\rho}
\nonumber \\
&\quad 
+
\frac{\theta}{8\pi^2} 
\epsilon^{\mu\nu\lambda\rho}
\partial_{\mu} a^{(i)}_{\nu} \partial_{\lambda}a^{(i)}_{\rho}
-e_i j^{(i)}_{\mu}A^{\mu},
\end{align}
for $i=1,\ldots \mathsf{p}$-th flavors, 
and 
\begin{align}
&\mathcal{L}^{(\mathsf{p}+1)}
=
-
\epsilon^{\mu\nu\lambda\rho}
\sum_{I=1}^{\mathsf{m}}
b^{I}_{\mu\nu} \partial_{\lambda}a^{I}_{\rho}
\nonumber \\
&\quad 
+
\frac{\theta}{8\pi^2} 
\epsilon^{\mu\nu\lambda\rho}
\sum_{I=1}^{\mathsf{m}}
\partial_{\mu}a^{I}_{\mu} \partial_{\lambda}a^{I}_{\rho}
-e_{\mathsf{p}+1} \sum_{I=1}^{\mathsf{m}} j^{I}_{\mu}A^{\mu}, 
\end{align}
for the $\mathsf{p}+1$-th flavor. 
Here,
$e=\sum_i e_i =1$, 
the parton densities 
are written in terms of 
the two-form gauge fields 
$
b^{(i)}_{\mu\nu}
$
and
$
b^{I}_{\mu\nu}
$
as
$j^{I,\mu}
= \epsilon^{\mu\nu\lambda\rho} \partial^{\ }_{\nu} b^I_{\lambda\rho}
$,
\emph{etc.}, 
and are subject to the constraint
\begin{align}
  \epsilon^{\mu\nu\lambda\rho} \partial^{\ }_{\nu}
  b^{(i)}_{\lambda\rho} 
=
\epsilon^{\mu\nu\lambda\rho} \partial^{\ }_{\nu}
b^{(\mathsf{p}+1)}_{\lambda\rho}
=
\sum_{I=1}^{\mathsf{m}}
\epsilon^{\mu\nu\lambda\rho} \partial^{\ }_{\nu}
b^{I}_{\lambda\rho},
\end{align}
($i=1,\ldots,\mathsf{p}$).
Solving the constraint,
following similar steps to those leading to
Eq.\ (\ref{parton lagrangian 2}), 
the resulting effective field theory is 
\begin{align}
\mathcal{L}
&=
-
\sum_{I}  
\epsilon^{\mu\nu\lambda\rho}
b^{I}_{\mu\nu} \partial_{\lambda}
\left(
a^{I}_{\rho}
+ \sum\nolimits_{i} a^{{(i)}}_{\rho}
\right)
\nonumber \\
&\quad
+
\frac{\theta}{8\pi^2}
\sum_{I}  
\epsilon^{\mu\nu\lambda\rho}
\partial_{\mu}a^{I}_{\nu} \partial_{\lambda}a^I_{\rho}
\nonumber \\
&\quad
+
\frac{\theta}{8\pi^2}
\sum_{i}  
\epsilon^{\mu\nu\lambda}
\partial_{\mu} a^{(i)}_{\nu} \partial_{\lambda}a^{(i)}_{\rho}
- e 
\sum_I
\epsilon^{\mu\nu\lambda\rho}
\partial_{\nu} 
b^I_{\lambda\rho}
A^{\mathrm{ex}}_{\mu}.
\end{align}

For a simple case
where $\mathsf{m}=1$ and $\mathsf{p}=
\mathsf{k}-1$,
introducing $b_{\mu\nu}:=b^{I=1}_{\mu\nu}$
and
labeling the $\mathsf{k}$ gauge fields 
as
$\alpha^{a=1,2,\ldots, \mathsf{k}}_{\mu}=
(
a^{(1)}_{\mu}, a^{(2)}_{\mu}, 
\ldots, 
a^{I=1}_{\mu})$,
the effective Lagrangian is given by 
\begin{align}
\mathcal{L}
&=
-
\epsilon^{\mu\nu\lambda\rho}
b_{\mu\nu} \partial_{\lambda}
\sum_{a=1}^\mathsf{k} \alpha^{a}_{\rho}
\nonumber \\
&\quad
+
\frac{\theta}{8\pi^2}
\sum_{a=1}^\mathsf{k}
\epsilon^{\mu\nu\lambda\rho}
\partial_{\mu}\alpha^{a}_{\nu} \partial_{\lambda} 
\alpha^a_{\rho}
- e 
\epsilon^{\mu\nu\lambda\rho}
\partial_{\nu} b_{\lambda\rho}
A^{\mathrm{ex}}_{\mu}.
\end{align}

As before in the case of $D=2+1$, 
we can eliminate $\alpha^a_{\mu}$ 
one by one. 
[The following steps should be compared 
with 
Eqs.\
(\ref{parton lagrangian reduced 1}
-
\ref{parton lagrangian reduced 2})
in the $D=2+1$ dimensional case.]
From
$
\delta S/\delta \alpha^a_{\mu}=
\delta S/\delta b_{\mu\nu}=0$,
\begin{align}
&
\epsilon^{\mu\nu\lambda\rho}
\partial_{\nu} b_{\lambda\rho}
-
\frac{1}{4\pi^2}
\epsilon^{\mu\nu\lambda\rho}
\partial_{\nu}
\left(
\theta \partial_{\lambda} \alpha^a_{\rho}
\right)
=0, 
\nonumber \\
&
\epsilon^{\mu\nu\lambda\sigma} \partial_{\lambda} 
\left(\sum\nolimits_a\alpha^a_{\sigma}- A^{\mathrm{ex}}_{\sigma}
\right)
=0.
\end{align}
If we assume $\theta=\mathrm{const.}$ and
neglect monopole configurations,
the first equation gives 
$\epsilon^{\mu\nu\lambda\rho}
\partial_{\nu} b_{\lambda\rho}=0$.
However, if we take into account monopoles, 
we claim, from the first equation
\begin{align}
b_{\lambda\rho}
=
\frac{1}{4\pi^2}
\theta \partial_{\lambda} \alpha^a_{\rho}, 
\quad
a = 1,\ldots, \mathsf{k}. 
\end{align}
Then $\alpha^1=\alpha^2=\cdots 
=\alpha^\mathsf{k} \equiv \alpha$, 
and hence the Lagrangian, after eliminating 
$\alpha^a_{\mu}$, is 
\begin{align}
\mathcal{L}
&=
-
\mathsf{k}
\epsilon^{\mu\nu\lambda\rho}
b_{\mu\nu} \partial_{\lambda}
\alpha_{\rho}
+
\frac{\mathsf{k}\theta}{8\pi^2}
\epsilon^{\mu\nu\lambda\rho}
\partial_{\mu}\alpha_{\nu} \partial_{\lambda} \alpha_{\rho}
\nonumber \\
&\quad 
- e 
\epsilon^{\mu\nu\lambda\rho}
\partial_{\nu} b_{\lambda\rho}
A^{\mathrm{ex}}_{\mu}.
\end{align}
Further integrating over $\alpha_{\mu}$
by using
$\delta S/\delta b_{\mu\nu}=0$,
\begin{align}
\epsilon^{\mu\nu\lambda\sigma} \partial_{\lambda} 
\left(\mathsf{k}\alpha_{\sigma}- A^{\mathrm{ex}}_{\sigma}
\right)
=0, 
\end{align}
we then arrive at
\begin{align}
\mathcal{L}
&=
\frac{ \theta}{8\pi^2 \mathsf{k}}
\epsilon^{\mu\nu\lambda\rho}
\partial_{\mu}A^{\mathrm{ex}}_{\nu} 
\partial_{\lambda} A^{\mathrm{ex}}_{\rho}. \label{eq:frac3dresponse}
\end{align}
[This step should be compared with
Eq.\ (\ref{3+1 axion free}).]

\subsection{Parton Construction for the Quantum Spin Hall Insulator in $D=2+1$} 
There are two natural ways to apply the parton construction for $D=2+1$ time-reversal invariant insulators: (i) create the necessary copies of the gauge fields in $D=4+1$ and then perform dimensional reduction twice (ii) perform dimensional reduction from a non-fractionalized theory in $D=4+1$ and then create replicas of the relevant fields. We will take the former approach. Our result, in fact, matches what would be found by taking Eq.\ (\ref{eq:frac3dresponse}) and simply performing dimensional reduction on $A^{\mathrm{ex}}_{\mu}.$

We begin with Eq.\ (\ref{eff FT for QSHE}) and add replicas for the $\phi, \psi, a_\mu, g_{\mu\nu}, u_{\mu\nu}$ and $v_\mu$ fields to find the Lagrangians
\begin{align}
\mathcal{L}^{(i)}
&=
-
\frac{\epsilon^{\mu \nu \lambda}}{2\pi}
(\phi^{(i)}-e_i\theta^{\mathrm{ex}})\partial_\mu g^{(i)}_{\nu\lambda}
\nonumber \\
&\quad
-
\frac{\epsilon^{\mu \nu \lambda}}{2\pi}(\psi^{(i)}-e_i\chi^{\mathrm{ex}})\partial_\mu u^{(i)}_{\nu\lambda}
\nonumber \\
&\quad
-
\frac{\epsilon^{\mu \nu \lambda}}{2\pi}
(a^{(i)}_{\mu}-e_i A^{\mathrm{ex}}_\mu) \partial_\nu v^{(i)}_{\lambda}
+
\frac{\epsilon^{\mu \nu \lambda}}{4\pi^2}
\phi^{(i)} 
\partial_\mu \psi^{(i)}
\partial_{\nu}
a^{(i)}_{\lambda},
\end{align}
where $i=1,\ldots, \mathsf{p}$.
Note that we have crucially included the $e_i$
charges in front of the scalar fields $\theta^{\mathrm{ex}}, \chi^{\mathrm{ex}}$ as they arise from the dimensionally reduced 
$A^{\mathrm{ex}}_{w}, A^{\mathrm{ex}}_{z}$ fields respectively and would enter the $D=4+1$ Lagrangian with the corresponding charge.
For the $\mathsf{p}+1$-st flavor we have 
\begin{align}
\mathcal{L}^{(\mathsf{p}+1)}
&=
-
\sum_{I=1}^{\mathsf{m}}
\frac{\epsilon^{\mu \nu \lambda}}{2\pi}(\phi^{I}-e_I\theta^{\mathrm{ex}})\partial_\mu g^{I}_{\nu\lambda}
\nonumber \\
&\quad 
-
\sum_{I=1}^{\mathsf{m}}
\frac{\epsilon^{\mu \nu \lambda}}{2\pi}
(\psi^{I}-e_I\chi^{\mathrm{ex}})\partial_\mu u^{I}_{\nu\lambda}
\nonumber \\
&\quad
-
\sum_{I=1}^{\mathsf{m}}\epsilon^{\mu \nu \lambda}
(a^{I}_{\mu}-e_{I}A^{\mathrm{ex}}_\mu) \partial_\nu v^{I}_{\lambda}
\nonumber \\
&\quad 
+
\sum_{I=1}^{\mathsf{m}}
\frac{\epsilon^{\mu \nu \lambda}}{4\pi^2}
\phi^{I} 
\partial_\mu \psi^{I}
\partial_{\nu}
a^{I}_{\lambda}. 
\label{parton QSHE p1}
\end{align}
For the simple case when
$\mathsf{m}=1, \mathsf{p}=\mathsf{k}-1$
the constraint for the parton densities identifies all of the $v^{(i)}_{\mu}\equiv v^{i}, g_{\mu\nu}^{(i)}\equiv g_{\mu\nu}$ and $u^{(i)}_{\mu\nu}=u_{\mu\nu}.$ We can then use the equation of motion
\begin{equation}
\frac{\delta S}{\delta a^{(i)}_{\mu}}
=
\epsilon^{\mu\nu\lambda}\left[
\frac{1}{4\pi^2}\partial_\nu (\phi^{(i)}\partial_\lambda \psi^{(i)})-\frac{1}{2\pi}\partial_\nu v_\lambda\right]=0
\end{equation}
to eliminate $a_{\mu}^{(i)}$ and arrive at
 \begin{align}
\mathcal{L}
&=
\frac{\epsilon^{\mu \nu \lambda}}{2\pi}
\left[
  \Big(
    \theta^{\mathrm{ex}}
    -\sum\nolimits_{i=1}^{\mathsf{k}}\phi^{(i)}
  \Big)\partial_\mu g_{\nu\lambda}+
\right.
\nonumber \\
&\quad
\qquad 
+
\left.
  \Big(\chi^{\mathrm{ex}}
    -\sum\nolimits_{i=1}^{\mathsf{k}}
\psi^{(i)}
\Big)\partial_\mu u_{\nu\lambda}\right]
\nonumber\\
&\quad +
\frac{\mathsf{k}}{4\pi^2}
\epsilon^{\mu \nu \lambda}
A^{\mathrm{ex}}_{\mu}\partial_{\nu}\Omega_\lambda, 
\end{align}
where 
$\Omega_\lambda =
\phi^{(i)}\partial_\lambda \psi^{(i)}$ which is the same for all $i$ from the equation of motion 
$\delta S/\delta a^{(i)}_{\mu}=0.$
Using
$\delta S/\delta \phi^{(i)}=0$
and
$\delta S/\delta \psi^{(i)}=0$
we can show that $\psi^{(i)}\equiv \psi$ and $\phi^{(i)}\equiv \phi$ are identical for all $i$ respectively. Eliminating $\psi$ and $\phi$ we finally arrive at
\begin{equation}
  \mathcal{L}=\frac{1}{4\pi^2 \mathsf{k}}
  \epsilon^{\mu\nu\lambda}
  A^{\mathrm{ex}}_{\mu}\partial_{\nu}
  (\theta^{\mathrm{ex}}\partial_\lambda \chi^{\mathrm{ex}}).
\end{equation}\noindent Physically this term implies that on the edge of a fractional quantum spin Hall system there will be fractional multiples of $e/2$ charge \emph{i.e.} 
$Q=e/(2 \mathsf{k})$ on an anti-phase magnetic domain wall.

\section{Discussion}
\label{sec: Discussion}

An effective Chern-Simons field theory approach
is one of the most successful theoretical frameworks
of the $D=2+1$ fractional quantum Hall effect.
In this paper, we aimed to extend this type of hydrodynamic formulation
to a broader class of non-interacting,
as well as interacting topological insulators,
in arbitrary dimensions with a suitable set of discrete symmetries.

We close with a few relevant comments.
First,
our approach relies crucially on the presence of a U(1) gauge symmetry.
While we have focused on topological insulators
with the electromagnetic charge U(1) symmetry,
there are topological phases that preserve non-Abelian symmetry
(such as spin-singlet topological superconductors in $D=2+1$ and $3+1$
that preserve spin SU(2) rotation symmetry).
Our approach can easily be extended to such situations
(see Appendix \ref{non-Abelian functional bosonization}).
There are also various topological phases that do not
conserve any quantum numbers, except possibly energy and momentum.
In particular, these include topological superconductors
in symmetry class D ($D=2$) and DIII ($D=3$).
For such topological superconductors,
Ref.\ \onlinecite{Hansson2011}
proposed a BF type topological field theory with
fermionic degrees of freedom.

Second, we note that in various situations, it may be useful to initially consider more
 U(1) charges than are actually conserved.
For example,
in the quantum spin Hall effect,
while the electromagnetic U(1) symmetry is strictly conserved,
spin rotation symmetry is not.
It is, however, still a useful starting point
to consider a system with U(1) symmetries
associated with both charge and spin rotations around, for example, the
$z$-axis.
The system is then invariant under an expanded $\mathrm{U}(1)\times \mathrm{U}(1)$
symmetry.
Functional bosonization in this case then gives rise to
a BF theory with Chern-Simons terms.
This is in fact the usual approach and gives rise to 
doubled Chern-Simons theories with
$\mathrm{U}(1)\times \mathrm{U}(1)$ symmetry
(more generally, with $\mathrm{U}(1)^N\times \mathrm{U}(1)^N$ symmetry).
Ultimately, however, the quantum spin Hall effect does not
depend crucially on the presence of the additional U(1) spin symmetry;
its stability is guaranteed by a $\mathbb{Z}_2$ topological invariant,
which has nothing to do with the presence of
the $S_z$ spin rotation symmetry.
Starting from the BF theory
with $\mathrm{U}(1)\times \mathrm{U}(1)$ symmetries,
the ``unwanted'' $S_z$ conservation can be broken
by introducing monopole processes or possibly the Higgs mechanism.
The former was discussed for the BF theory describing a BCS
superconductor in Ref.\ \onlinecite{Hansson04},
which has a $\mathbb{Z}_2$ topological order.
It would be interesting to see if one started from a 
$\mathrm{U}(1)\times \mathrm{U}(1)$ theory and
implemented either symmetry breaking mechanism
if one would generate a hydrodynamic field theory that
is equivalent with what we constructed in this article
by only using the $\mathrm{U}(1)$ charge symmetry.

Finally, due to the lack of microscopic realizations
of fractional topological insulators (in particular in $D>2+1$),
the usefulness of our effective field theory approach 
is not yet entirely clear.
It is, however, interesting to speculate on the
possible mechanism of such states.
In the functional bosonization scheme,
we need to ``raise the level'' of the BF term,
as it is this term that controls the ground state degeneracy.
In this paper, we have explored the parton construction.
Another possible way to raising the level would be
to use a Higgs field.
This mechanism was discussed in the fractional quantum Hall effect,
in particular, in non-Abelian  fractional quantum Hall states.
Together with functional bosonization,
this method may allow us to discuss the same kinds of
fractional states obtained by the parton construction,
as well as different kinds of states.
\cite{Fradkin1999, Fradkin1997}

\begin{acknowledgments}

We thank fruitful interactions with
Maissam Barkeshli, 
Andrei Bernevig,
Thomas Faulkner,
Matthew Fisher, 
Hans Hansson,
Ken Nomura,
Shivaji Sondhi,
T. Senthil,
Xiao-Liang Qi,
and
Ashvin Vishwanath.
This work was supported in part by U.S. DOE under Award DE-FG02-07ER46453 (TLH),
and by the NSF, under grant DMR-1064319 (EF) 
at the University of Illinois.
\end{acknowledgments}

\appendix

\section{Non-Abelian functional bosonization}
\label{non-Abelian functional bosonization}

Here, we describe the functional bosonization for non-Abelian currents. 
We start from the generating functional
$Z[A^{\mathrm{ex}}_{\mu}]$
for the correlation functions of non-Abelian currents, 
where $A^{\mathrm{ex}}_{\mu}$ is a Lie-algebra valued external field. 
We rewrite $Z[A^{\mathrm{ex}}_{\mu}]$ as
\begin{align}
Z[A^{\mathrm{ex}}_{\mu}] = 
\int \mathcal{D}[a_{\mu}]
\prod_x \prod_{\mu}\delta[a_{\mu}-A^{\mathrm{ex}}_{\mu}]
Z[a_{\mu}]. 
\end{align}
The delta functional can be written as
\begin{align}
&\quad 
\prod_x \prod_{\mu}\delta[a_{\mu}-A^{\mathrm{ex}}_{\mu}]
\nonumber \\
&=
\Delta_{\mathrm{FP}}[a]
\prod_x
\prod^{\mu<\nu<\lambda\cdots}_{\mu,\nu,\lambda,\ldots}
\epsilon_{\mu\nu\lambda\cdots\alpha\beta}
\prod_a
\delta[f^a_{\alpha\beta}[a]-f^a_{\alpha\beta}[A^{\mathrm{ex}}]], 
\label{delta func}
\end{align}
where
$f_{\alpha\beta}[a]=\partial_{\alpha} a_{\beta}-\partial_{\beta} a_{\alpha}
+[a_{\alpha},a_{\beta}]$,
and 
$\prod_{a}$ runs over the Lie algebra index, 
and 
$\prod_{\mu,\nu}$ runs over $n=D(D-1)/2$ independent directions. 
The Jacobian $\Delta_{\mathrm{FP}}[a]$ 
can be expressed in terms of a functional integral as follows.
By integrating over $A^{\mathrm{ex}}_{\mu}$ in Eq.\ (\ref{delta func}),
\begin{align}
1&=
\int\mathcal{D}[A_{\mu}]
\prod_x \prod_{\mu}\delta[a_{\mu}-A_{\mu}]
\nonumber \\
&=
\Delta_{\mathrm{FP}}[a]
\int\mathcal{D}[A_{\mu}]
\nonumber \\
&\qquad 
\times 
\prod_x 
\prod^{\mu<\nu<\rho\cdots}_{\mu,\nu,\rho,\ldots}
\epsilon_{\mu\nu\rho\cdots}
\prod_a
\delta[f^a_{\alpha\beta}[a]-f^a_{\alpha\beta}[A]]. 
\end{align}
(Here we denote $A^{\mathrm{ex}}_{\mu}=A_{\mu}$ to lighten notations).
Consider 
\begin{align}
(*):=
\int\mathcal{D}[A]
\prod_x \prod_{\alpha,\beta}^{\alpha<\beta}
\delta
\left(
f^a_{\alpha\beta}[a]-f^a_{\alpha\beta}[A]
\right).
\end{align}
We expand  
$f_{\alpha\beta}[A]$
as
\begin{align}
f_{\alpha\beta}[A]
=
f_{\alpha\beta}[a]
+ 
D_{\alpha}[a] \delta A_{\beta}
-
D_{\beta}[a] \delta A_{\alpha}
\end{align}
where $A=a + \delta A$.
Then, 
\begin{align}
(*)&=
\int \mathcal{D}[\delta A_{\mu}]
\prod_x 
\prod^{\alpha<\beta}_{\alpha,\beta}
\delta(
D^a_{\alpha}[a] \delta A_{\beta}
-
D^a_{\beta}[a] \delta A_{\alpha}
). 
\end{align}
This can be rewritten 
with a bosonic auxiliary field $\beta_{\alpha\beta}$ as 
\begin{align}
(*)
&=
\int \mathcal{D}[\beta_{\alpha\beta}, \delta A_{\mu}]
\nonumber \\
&\quad
\times
\exp
{i} 
\int d^Dx\,
\sum_{\alpha<\beta}
\beta^a_{\alpha\beta}
\left(
D^a_{\alpha}[a] \delta A_{\beta}
-
D^a_{\beta}[a] \delta A_{\alpha}
\right), 
\end{align}
where 
$
\beta_{\alpha\beta}
=
-
\beta_{\beta\alpha}
$
and 
$\beta_{\alpha\beta}=0$
for $\beta=\alpha$.
Thus
\begin{align}
\Delta_{\mathrm{FP}}[a]^{-1}
&=
\int \mathcal{D}[\beta_{\alpha\beta}, \delta A_{\mu}]
\exp
{i} \int d^Dx\,
\beta^a_{\alpha\beta}
D^a_{\alpha}[a] \delta A_{\beta}. 
\end{align}
The functional determinant can then be 
written, in terms of fermionic ghosts 
$\bar{c}_{\alpha\beta}$ and $c_{\mu}$, 
\begin{align}
\Delta_{\mathrm{FP}}[a]
&=
\int \mathcal{D}[\bar{c}_{\alpha\beta}, c_{\mu}]
\exp
{i} 
\int d^Dx\,
\bar{c}^a_{\alpha\beta}
D_{\alpha}[a] c^a_{\beta}. 
\end{align}
Then, the generating functional can be written as
\begin{align}
&Z[A^{\mathrm{ex}}] = \int \mathcal{D}[a,b,\bar{c},c]
\exp {i} S,
\nonumber \\
&S=
\int d^Dx\, \mathrm{tr}\,
\Big[
\bar{c}_{\mu\nu\cdots}
\epsilon^{\mu\nu\cdots\alpha\beta}
D_{\alpha}[a] c_{\beta}
\nonumber \\
&\quad\quad  
+
b_{\mu\nu\cdots}
\epsilon^{\mu\nu\cdots\alpha\beta}
\left(
f_{\alpha\beta}[a]
-
f_{\alpha\beta}[A^{\mathrm{ex}}]
\right)
+
L[a]
\Big], 
\end{align}
where we introduced $L[a]$ by
$Z[a] =\exp {i} \int d^Dx\, \mathrm{tr}\, L[a]$. 

Let us now introduce auxiliary fields
\begin{align}
S&=
\int d^Dx\, \mathrm{tr}\,
\Big[
\bar{c}_{\mu\nu\cdots}
\epsilon^{\mu\nu\cdots\alpha\beta}
D_{\alpha}[a] c_{\beta}
\nonumber \\
&\quad 
+
b_{\mu\nu\cdots}
\epsilon^{\mu\nu\cdots\alpha\beta}
\left(
f_{\alpha\beta}[a]
-
f_{\alpha\beta}[A^{\mathrm{ex}}]
\right)
\nonumber \\
&\quad
+
L[a-h]
-{i}
(l h^{\mu}h_{\mu} 
- 2 \bar{\chi} h^{\mu} c_{\mu})
\Big].
\end{align}
Here,
$l$ and $\bar{\chi}$
are a bosonic and fermionic scalar, respectively,
and $h_{\mu}$ is a bosonic, Lie-algebra valued field,
$h_{\mu} = h^a_{\mu}t_a$. 
The integration over $l$ sets
$h^{\mu}h_{\mu}=0 
\Rightarrow
h_{\mu}=0$ 
and then we go back to the original action.
The action is invariant under the following BRST transformation:
\begin{align}
\delta a_{\mu}
&=
c_{\mu},
\quad
\delta h_{\mu}
=
c_{\mu},
\nonumber \\
\delta \bar{\chi} 
&= l,
\quad
\delta l
= 0,
\nonumber \\
\delta c_{\mu}
&=
0, 
\quad 
\delta \bar{c}_{\mu\nu\cdots}
= - 2b_{\mu\nu\cdots}, 
\nonumber \\
\delta b_{\mu\nu\cdots}
&=0. 
\end{align}
It is possible to write the action 
as
\begin{align}
{i}S 
&= \delta G,
\nonumber \\
\mbox{where}
\quad 
G
&=
-
\frac{{i}}{2}
\int d^Dx\, \mathrm{tr}\,
\left[
\bar{c}_{\mu\nu\cdots}
\epsilon^{\mu\nu\cdots\alpha\beta}
f_{\alpha\beta}[a]
\right]. 
\end{align}

\section{Quantization of the BF theory with
the Chern-Simons and axion term}

From the bosonized action, we can read off,
\textit{e.g.,}
the current-current correlation functions.
If we take the BF-CS theory (\ref{bf-cs theory}),
however, we do not have charge fluctuations
in the bulk of the system;
with only terms of topological origin (such as BF and CS terms),
there is no dynamics.
This situation corresponds to the limit
where the band gap (mass) is taken to be infinitely large.  
To see charge fluctuations,
we need either to
have other terms than BF and CS terms,
or to introduce a boundary to a system.
The former means we
keep the band gap finite,
or consider electron-electron interactions. 
Let us thus consider
the Lagrangian
with the Maxwell term
\begin{align}
\mathcal{L} 
&=
-
\frac{2\mathsf{k}}{4\pi} 
b_{\mu} \epsilon^{\mu\nu\lambda} \partial_{\nu} 
(a_{\lambda}-A_{\lambda})
+
\frac{\mathsf{Ch}}{4\pi} 
\epsilon^{\mu\nu\lambda}
a_{\mu}\partial_{\nu} a_{\lambda}
\nonumber \\
& \quad - \frac{1}{4\pi g^2} F_{\mu\nu} F^{\mu\nu},
\label{bf-cs maxwell theory}
\end{align}
where 
$g^2$ is a coupling constant, 
$A_{\mu}$ is the {\it dynamical} electromagnetic U(1)
gauge field and
$F_{\mu\nu}= \partial_{\mu} A_{\nu} - \partial_{\nu}A_{\mu}
$
is the field strength.
[$A_{\mu}$ is not a static source as in Eq.\ (\ref{bf-cs theory})].
Given the equation of motion
$\delta S/\delta b_{\mu}=0 \Rightarrow F_{\mu\nu}[A]=f_{\mu\nu}[a]$, 
the Maxwell term 
$- (4\pi g^2)^{-1} F_{\mu\nu} F^{\mu\nu}$
gives rise to the Maxwell term for $a_{\mu}$,
$- (4\pi g^2)^{-1} f_{\mu\nu}f^{\mu\nu}$. Alternatively,
integrating over $A_{\mu}$ gives rise to
a current-current interaction
$\propto j^{\mu}(x) K_{\mu\nu}(x-y) j^{\nu}(y)$
where according to the bosonization rule,
$j^{\mu}  \propto
\epsilon^{\mu\nu\lambda} \partial_{\nu} b_{\lambda}.
$

The BF-CS-Maxwell theory (\ref{bf-cs maxwell theory})
can be canonically quantized by going through the Dirac
quantization procedure.
Some non-zero canonical commutation relations are given by
\begin{align}
&
\big[
  b_i (\vec{x}), b_j(\vec{x}')
\big]
=
\frac{-{i} 2\pi \mathsf{Ch} }{\mathsf{k}^2}
\epsilon^{ij}
\delta^{(2)}(\vec{x}-\vec{x}'),
\nonumber \\
&
\big[
  a_i(\vec{x}), \pi^{j}_a(\vec{x}')
\big]
=
\big[
  A_i(\vec{x}), \pi^j_A(\vec{x}')
\big]
=
{i}\delta^j_i \delta^{(2)}(\vec{x}-\vec{x}'),
\end{align}
where 
$\vec{x}=(x^1,x^2)$ is the spatial coordinates, 
$i,j$ run over the spatial
components of the coordinates
$i,j=x,y$, and 
$\pi^i_a$ and $\pi^i_A$ are the canonical
momentum for $a_i$ and $A_i$ and are given by
\begin{align}
  \pi^i_a
  &= 
  -\frac{2\mathsf{k}}{4\pi}
  \epsilon^{ij} b_j
  +
  \frac{\mathsf{Ch}}{4\pi}
  \epsilon^{ij}a_j, 
\nonumber \\
  \pi^i_A
  &= 
  +\frac{2\mathsf{k}}{4\pi}
  \epsilon^{ij} b_j
  +
  \frac{1}{\pi g^2}
  E_i, 
\end{align}
where $E_i = \partial_0 A_i - \partial_i A_0$. 
The canonical commutation relations between current
operators are given by
\begin{align}
  \left[ j^0(\vec{x}), j^0(\vec{x}')\right]
  &=0,
\nonumber \\
  \left[ j^0(\vec{x}), j^i(\vec{x}')\right]
  &=
\frac{ -{i} \mathsf{Ch}^2 g^2}{4\pi} \partial_i
\delta^{(2)}(\vec{x}-\vec{x}'),
\nonumber \\
  \left[ j^i(\vec{x}), j^j(\vec{x}')\right]
  &=
\frac{ -{i} \mathsf{Ch}^3 g^2}{8 \pi } 
\epsilon^{ij} 
\delta^{(2)}(\vec{x}-\vec{x}'). 
\label{current alg. 2+1 d}
\end{align}

Similarly,
in $D=(3+1)$ dimensions,
one can use the bosonized action
(\ref{3+1 action}) to read off the current-current commutation
relations in the bulk, once we allow charges to fluctuate,
by adding an Maxwell term, say.  
Let us consider the Lagrangian, 
\begin{align}
\mathcal{L}
&=
-
b_{\mu\nu} 
\epsilon^{\mu\nu\lambda\rho} \partial_{\lambda} 
(a_{\rho}-A_{\rho})
\nonumber \\
&\qquad 
+
\frac{\theta(x)}{8\pi^2} 
\epsilon^{\mu\nu\lambda\rho}
\partial_{\mu} a_{\nu}\partial_{\lambda} a_{\rho}
-
\frac{1}{4\pi g^2}
F_{\mu\nu} F^{\mu\nu},
\label{bf-axion maxwell theory}
\end{align}
where as before $A_{\mu}$ is a dynamical U(1) gauge field.

The BF-axion-Maxwell theory (\ref{bf-axion maxwell theory})
can be canonically quantized by going through the Dirac
quantization procedure.
As one may infer from the (half-integral) quantum Hall effect
at the surface of $D=3+1$ dimensional topological insulators,
the current-current commutators are ``anomalous''
only at the location where $\theta(x)$ changes, 
{\it i.e.}, at an interface between topologically trivial
and topologically non-trivial insulators. 
For simplicity, we assume $\theta(x)$ depends
only on the $x^3$ ($z$) component of the spatial coordinates.  
Some non-zero canonical commutation relations are given by
\begin{align}
&
\big[
  b_{m3} (\vec{x}), b_{n3}(\vec{x}')
\big]
=
\frac{-{i} (\partial_3 \theta) }{16\pi^2}
\epsilon^{mn}
\delta^{(3)}(\vec{x}-\vec{x}'),
\nonumber \\
&
\big[
  a_m(\vec{x}), \pi^{n}_a(\vec{x}')
\big]
=
\big[
  A_m(\vec{x}), \pi^n_A(\vec{x}')
\big]
=
{i}\delta^n_m \delta^{(3)}(\vec{x}-\vec{x}'),
\end{align}
where $m,n$ run over the spatial
components of the coordinates
$m,n=x,y$, and 
$\pi^i_a$ and $\pi^i_A$ are the canonical
momentum for $a_i$ and $A_i$ and are given by
\begin{align}
  \pi^i_a
  &= 
  -
  \epsilon^{ijk} b_{jk}
  +
  \frac{\theta}{4\pi^2}
  \epsilon^{ijk} \partial_j a_k, 
\nonumber \\
  \pi^i_A
  &= 
  +
  \epsilon^{ijk} b_{jk}
  +
  \frac{1}{\pi g^2}
  E_i. 
\end{align}
The canonical commutation relations between current
operators are given by
\begin{align}
  \left[ j^0(\vec{x}), j^0(\vec{x}')\right]
  &=0,
\nonumber \\
  \left[ j^0(\vec{x}), j^m(\vec{x}')\right]
  &=
\frac{ -{i} (\partial_3 \theta)^2 g^2}{16\pi^3 } \partial_m
\delta^{(3)}(\vec{x}-\vec{x}'),
\nonumber \\
  \left[ j^m(\vec{x}), j^n(\vec{x}')\right]
  &=
\frac{ -{i} (\partial_3 \theta)^3 g^2}{64 \pi^4 } 
\epsilon^{mn} 
\delta^{(3)}(\vec{x}-\vec{x}'). 
\end{align}
This is exactly the same as
the quantum Hall effect
in the topological insulator in $2+1$ dimensions,
Eq.\ (\ref{current alg. 2+1 d}),
but with the
integer $\mathsf{Ch}$ replaced by
$\partial_3\theta$.

\newpage


\begin{thebibliography}{99}


\bibitem{review_QHE}
{\it The Quantum Hall Effect},
edited by R. E. Prange and S. M. Girvin (Springer, New York, 1987).

\bibitem{review_TIa}
M. Z. Hasan, and C. L. Kane,
Rev. Mod. Phys. \textbf{82}, 3045 (2010).

\bibitem{review_TIb}
X.-L. Qi, and S.-C. Zhang,
Rev. Mod. Phys. \textbf{83}, 1057 (2011).


\bibitem{KaneMele}
C.\ L.\ Kane and E.\ J.\ Mele,
Phys.\ Rev.\ Lett. \textbf{95}, 146802 (2005);
C.\ L.\ Kane and E.\ J.\ Mele,
Phys.\ Rev.\ Lett. \textbf{95}, 226801 (2005).

\bibitem{Roy06}
R.\ Roy,
Phys.\ Rev.\ B \textbf{79}, 195321 (2009).

\bibitem{Bernevig05}
B.\ A.\ Bernevig and S.-C. Zhang,
Phys.\ Rev.\ Lett. \textbf{96}, 106802 (2006).


\bibitem{Moore06}
J.\ E.\ Moore and L.\ Balents,
Phys.\ Rev.\ B \textbf{75}, 121306(R) (2007).

\bibitem{Roy3d}
R.\ Roy,
Phys.\ Rev.\ B \textbf{79}, 195322 (2009).

R. Roy, Phys. Rev. B 79, 195321 (May 2009)
\bibitem{Fu06_3Da}
L.\ Fu, C.\ L.\ Kane, and E.\ J.\ Mele,
Phys.\ Rev.\ Lett. \textbf{98}, 106803 (2007).

\bibitem{Fu06_3Db}
L.\ Fu and C.\ L.\ Kane,
Phys.\ Rev.\ B \textbf{76}, 045302 (2007).

\bibitem{Qi_Taylor_Zhang2008}
X.-L.\ Qi, T.\ Hughes, and S.-C.\ Zhang,
Phys.\ Rev.\ B \textbf{78}, 195424 (2008).


\bibitem{3HeB}
See,
Y. Wada \textit{et al.},
Phys.\ Rev.\ B \textbf{78}, 214516 (2008),
S. Murakawa \textit{et al.},
Phys.\ Rev.\ Lett.\ \textbf{103}, 155301 (2009),
S.\ Murakawa {\it et al.},
J.\ Phys.\ Soc.\ Jpn.\ \textbf{80}, 013602 (2011),
and references therein.


\bibitem{Schnyder2008}
A. P. Schnyder, S. Ryu, A. Furusaki, and A. W. W. Ludwig,
Phys. Rev. B \textbf{78}, 195125 (2008).

\bibitem{SRFLnewJphys}
S. Ryu, A. Schnyder, A. Furusaki and A. W. W. Ludwig,
New J. Phys.
\textbf{12},
065010
(2010).

\bibitem{Kitaev2009}
A.\ Yu Kitaev,
\textit{AIP Conf.\ Proc.} \textbf{1134}, 22 (2009).


\bibitem{Essin08}
A.\ M.\ Essin, J.\ E.\ Moore, and D.\ Vanderbilt,
Phys.\ Rev.\ Lett.\ \textbf{102}, 146805 (2009).


\bibitem{Ryu2011}
S. Ryu, J. E. Moore, and A. W. W. Ludwig,
Phys. Rev. B \textbf{85},
045104 (2012).

\bibitem{Wang2011}
Z.\ Wang,
X.-L.\ Qi,
and
S.-C.\ Zhang,
Phys.\ Rev.\ B \textbf{84},  014527 (2011). 


\bibitem{NomuraStreda2011}
K.\ Nomura, 
S.\ Ryu,
A.\ Furusaki,
and 
N.\ Nagaosa,
Phys.\ Rev.\ Lett.\ \textbf{108} (2012) 026802.

\bibitem{niu-thouless-wu}
Q. Niu, D. J. Thouless and Y.-S. Wu, Phys. Rev. B {\bf 31}, 3372 (1985).

\bibitem{wen-niu}
X.-G. Wen and Q. Niu, Phys.\ Rev.\ B {\bf 41}, 9377 (1990).


\bibitem{wen-zee}
X.-G. Wen and A. Zee, Phys.\ Rev.\ B {\bf 46}, 2290 (1992).


%

\bibitem{Wen1995}
X.-G. Wen, 
Adv. Phys. \textbf{44}, 405 (1995).


\bibitem{ChoMoore2010}
G. Y. Cho and J. E. Moore, 
Annals Phys. \textbf{326} 1515-1535 (2011).


\bibitem{Nikolic}
P. Nikolic, 
\texttt{arXiv:1108.5388};
P. Nikolic, T. Duric and Z. Tesanovic, 
\texttt{arXiv:1109.0017};
P. Nikolic, 
\texttt{arXiv:1206.1055}.




\bibitem{Vishwanath2012}
A. Vishwanath, and T. Senthil,
\texttt{arXiv:1209.3058}.


\bibitem{Diamantini2012}
M. C. Diamantini, and C. A. Trugenberger,
\texttt{arXiv:1112.3281}. 

\bibitem{ZhangHanssonKivelson1989}
S. C. Zhang, T. H. Hansson, and S. Kivelson, 
Phys.\ Rev.\ Lett.\ \textbf{62}, 82 (1989).


\bibitem{Jain89a}
J.\ K.\ Jain, 
Phys.\ Rev.\ Lett.\ \textbf{63}, 199 (1989).

\bibitem{Jain89b}
J.\ K.\ Jain, 
Phys. Rev. B \textbf{40}, 8079 (1989). 

\bibitem{LopezFradkin1991}
A. L\'opez and E. Fradkin, Phys. Rev. B {\bf 44}, 5246 (1991).

\bibitem{FradkinBook}
E. Fradkin, {\it Field Theories of Condensed Matter Physics}, 2nd edition, Cambridge University Press (Cambridge, UK) (2012).

\bibitem{Blok1990a}
B.\ Blok and X.-G. Wen,
Phys.\ Rev.\ B, \textbf{42} 8133 (1990).

\bibitem{Blok1990b}
B.\ Blok and X.-G. Wen,
Phys.\ Rev.\ B, \textbf{42} 8145 (1990).



\bibitem{MattisLieb1965}
D. C. Mattis and 
E. H. Lieb, 
J. Math. Phys. \textbf{6}, 304 (1965). 

\bibitem{LutherPeschel1974}
A. Luther and
I. Peschel,
Phys. Rev. B \textbf{9}, 2911 (1974). 

\bibitem{LutherEmery1974}
A. Luther and V. J. Emery, Phys. Rev. Lett. {\bf 33}, 589 (1974).

\bibitem{Coleman1975}
S.\ Coleman, 
Phys. Rev. D \textbf{11}, 2088 (1975). 

\bibitem{Mandelstam1975}
S. Mandelstam, Phys. Rev. D {\bf 11}, 3026 (1975).



\bibitem{Haldane1981}
F. D. M. Haldane, J. Phys. C: Solid State Phys. {\bf 14}, 2585 (1981)

\bibitem{Gamboa1984}
R. E. Gamboa Sarav{\'\i}, M. A. Muschietti, F. A. Schaposnik, and J. E. Solomin,
Ann. Phys {\bf 157}, 360 (1984), and references therein.

\bibitem{Polyakov-Wiegmann}
A. M. Polyakov and P. B. Wiegmann, Phys. Lett. B {\bf 131}, 121 (1983); {\em ibid} {\bf 141}, 223 (1984).


\bibitem{Stone1994}
{\it Bosonization},  
edited by M. Stone, 
World Scientific (Singapore), (1994). 

\bibitem{Gogolin1998}
A. O. Gogolin, A. A. Nersesyan, and A. M. Tsvelik, {\em Bosonization and Strongly Correlated Systems}, Cambridge University Press (Cambridge, UK) (1998). 

\bibitem{Witten1984}
E. Witten, Comm. Math. Phys. {\bf 92}, 455 (1984).

\bibitem{Fujikawa}
K. Fujikawa, Phys. Rev. Lett. {\bf 42}, 1195 (1979); {\it ibid} Phys. Rev. D {\bf 21}, 2848 (1980).

\bibitem{Naon1985}
C. M. Na\'on, Phys. Rev. D {\bf 31}, 2035 (1985).

\bibitem{comment-tomography}
There have been many attempts at formulating a fermion-bose mapping for massless relativistic fermions in $D>2$ space-time dimensions. A notable example is the ``tomographic'' bosonization of Luther\cite{Luther-tomography} whose bosonic action is non-local. However, a variant of this approach has been used to bosonize dense fermi systems as a theory of a fluctuating Fermi surface.\cite{bosonizedFS}

\bibitem{Luther-tomography}
A. Luther, Phys. Rev. B {\bf 19}, 320 (1979).


\bibitem{bosonizedFS}
F. D. M. Haldane, Proceedings of the International School of Physics ``Enrico Fermi,'' course 121, Varenna, 1992, J. R. Schrieffer and R. Broglia editors,  North-Holland (NY) (1994); A. Houghton and J. B. Marston, Phys. Rev. B {\bf 48}, 7790 (1993); A. Houghton, H. J. Kwon and J. B. Marston, Adv. Phys. {\bf 49}, 141 (2000); A. H. Castro Neto and E. Fradkin, Phys. Rev. Lett {\bf 72}, 1393 (1994); {\em ibid} Phys. Rev. B {\bf 49}, 10877 (1994); {\it ibid} {\bf 51}, 4084 (1995).

\bibitem{Fradkin1995}
E.\ Fradkin and F.\ A.\ Schaposnik, 
Phys.\ Lett.\ B \textbf{338}, 253 (1995).

\bibitem{Schaposnik1995}
F.\ A.\ Schaposnik, Phys.\ Lett.\ B \textbf{356}, 39 (1995).

\bibitem{Barci2000}
D. G. Barci, C. A. Linhares, A. F. De Queiroz, 
and J. F. Medeiros Neto, 
Int.\ J.\ Mod.\ Phys.\ A \textbf{15}, 4655 (2000).

\bibitem{BarciOxman2000}
D. G. Barci and L. E. Oxman, 
Nucl. Phys. B \textbf{580}, 721 (2000).

\bibitem{Shizuya2001}
K.\ Shizuya, Phys. Rev. B {\bf 63}, 245301 (2001).

\bibitem{Schaposnik2001}
F.\ A.\ Schaposnik, 
AIP Conf. Proc. \textbf{419}, 151 (1998); \texttt{hep-th/9705186}.

\bibitem{diamond09}
A. P. Schnyder, S. Ryu, and A. W. W. Ludwig, 
Phys.\ Rev.\ Lett.\ \textbf{102}, 196804 (2009). 

\bibitem{Altland-Zirnbauer}
A. Altland and M. R. Zirnbauer, Phys. Rev. B {\bf 55} 1142 (1997).

\bibitem{Jackiw-Schrieffer}
R. Jackiw and J. R. Schrieffer, Nucl. Phys. B {\bf 190}, 253 (1981); J. Goldstone and F. Wilczek, Phys. Rev. Lett. {\bf 47}, 986 (1981).


\bibitem{FidkowskiKitaev2010}
L.\ Fidkowski and A.\ Kitaev, 
Phys.\ Rev.\ B \textbf{81}, 134509
(2010).

\bibitem{FidkowskiKitaev2011}
L.\ Fidkowski and A.\ Kitaev, 
Phys.\ Rev.\ B \textbf{83}, 075103
(2011).



\bibitem{Lopez1999}
A. L\'opez and E. Fradkin, Phys. Rev. B {\bf 59}, 15323 (1999).

\bibitem{Hosur2009}
P.\ Hosur, S.\ Ryu, and A.\ Vishwanath, 
Phys.\ Rev.\ B, \textbf{81}, 045120 (2010).

\bibitem{Balachandran1993}
A. P. Balachandran, and P. Teotonio-Sobrinho,
Int.\ J.\ Mod.\ Phys.\ A \textbf{8}, 723-752 (1993).


\bibitem{Bergeron1995}
 M.\ Bergeron,
 G.\ W.\ Semenoff,
 and
 R. J. Szabo,
 Nucl.\ Phys.\ B \textbf{437}, 695-722 (1995).


\bibitem{Hansson04}
T.\ H.\ Hansson, 
Vadim Oganesyan,
and
S.\ L.\ Sondhi,
Annals of Physics \textbf{313}, 497 (2004).



\bibitem{LeeKivelson2003}
D.-H. Lee and S. A. Kivelson,
Phys.\ Rev.\ B \textbf{67}, 024506 (2003).


\bibitem{Shindou07}
R.\ Shindou,
K.\ Imura,
and
M.\ Ogata, Phys. Rev. B {\bf 74}, 245107 (2006).

\bibitem{Peskin1978}
M.\ E.\ Peskin, 
Annals of Physics \textbf{113}, 122 (1978).

\bibitem{Dasgupta1981}
C.\ Dasgupta and B.\ I.\ Halperin, 
Phys.\ Rev.\ Lett.\ \textbf{47}, 1556 (1981).


\bibitem{Fisher1989}
M.\ P.\ A.\ Fisher and D.\ -H.\ Lee, 
Phys.\ Rev.\ B \textbf{39}, 2756 (1989); 
D.\ -H.\ Lee and M.\ P.\ A.\ Fisher, 
Phys.\ Rev.\ Lett.\ \textbf{63}, 903 (1989).


\bibitem{Haldane1983}
F.\ D.\ M.\ Haldane, 
Phys.\ Rev.\ Lett.\ \textbf{51}, 605 (1983).


\bibitem{Fradkin1989}
E.\ Fradkin, 
Phys.\ Rev.\ Lett.\ \textbf{63}, 322 (1989). 


\bibitem{Murthy2011}
G.\ Murthy and R. Shankar,
\texttt{arXiv:1108.5501}.

\bibitem{Murthy2012}
G.\ Murthy and R. Shankar,
Phys.\ Rev.\ B \textbf{86}, 195146 (2012).



 
\bibitem{Szabo1998}
R. J. Szabo,
Nucl.\ Phys.\ B \textbf{531} 525,(1998).

\bibitem{Szabo1999}
R. J. Szabo,
Ann.\ Phys.\ \textbf{280}, 163 (2000). 


\bibitem{Witten1979}
E.\ Witten, 
Phys.\ Lett.\ B \textbf{86}, 283 (1979).

\bibitem{Qi2009}
X. -L. Qi, R. Li, J. Zang, S. -C. Zhang,
Science \textbf{323}, 1184 (2009).

\bibitem{Rosenberg2010}
G.\ Rosenberg and M.\ Franz,
Phys.\ Rev.\ B \textbf{82}, 035105 (2010).


\bibitem{Kleinert}
H.\ Kleinert,
{\it Multivalued Fields
in Condensed Matter, Electromagnetism, and Gravitation}, 
World Scientific (Singapore), 
(2008). 

\bibitem{Quevedo1997}
F.\ Quevedo, 
and
C.\ A.\ Trugenberger, 
Nucl.\ Phys.\ B, 
\textbf{501}, 143 (1997).







\bibitem{Bernevig2006}
B.\ A. Bernevig, and S. -C. Zhang,
Phys.\ Rev.\ Lett. \textbf{96}, 106802 (2006).

\bibitem{Levin2009}
M.\ Levin, and A.\ Stern,
Phys.\ Rev.\ Lett.\ \textbf{103}, 196803 (2009). 

\bibitem{Levin2012}
M.\ Levin, and A.\ Stern,
Phys.\ Rev.\ B \textbf{86}, 115131 (2012).

\bibitem{Neupert2011}
T.\ Neupert, L.\ Santos, S.\ Ryu, C.\ Chamon, and C.\ Mudry,
Phys.\ Rev.\ B \textbf{84}, 165107 (2011).

\bibitem{Santos2011}
L.\ Santos, T.\ Neupert, S.\ Ryu, C.\ Chamon, and C.\ Mudry,
Phys.\ Rev.\ B \textbf{84}, 165138 (2011).


\bibitem{Lu2012}
Y. -M. Lu, and A. Vishwanath,
Phys.\ Rev.\ B \textbf{86}, 125119 (2012).



\bibitem{LeeZhang1991}
D.-H. Lee and S.-C. Zhang, 
Phys. Rev. Lett. \textbf{66}, 1220 (1991).

\bibitem{Zhang1992}
S. -C. Zhang, Int. J. Mod. Phys. B \textbf{6}, 25 (1992).


\bibitem{ZeeBook}
A.\ Zee, {Quantum Hall Fluids}, in
{\it Field Theory, Topology and Condensed Matter Physics}, 
edited by H. B.
Geyer (Springer, Berlin, 1995).


\bibitem{Wen1990}
X.\ -G.\ Wen, Int. J. Mod. Phys. B \textbf{4}, 239 (1990).

\bibitem{Wen1992}
X.\ -G.\ Wen, Int. J. Mod. Phys. B \textbf{6}, 1711 (1992).

\bibitem{WenBook}
X.\ -G.\ Wen, 
{\it Quantum Field Theory of Many-Body Systems}, 
Oxford University Press,
(2004).






\bibitem{Neupert11}
T. Neupert, L. Santos, C. Chamon, and C. Mudry,
Phys.\ Rev.\ Lett.\ \textbf{106}, 236804 (2011).


\bibitem{Sheng11}
D.\ N.\ Sheng, Z.\ Gu, K.\ Sun, and L.\ Sheng, 
Nature Communications \textbf{2}, 389 (2011).


\bibitem{Wang11}
Y.\ -F.\ Wang, Z.\ -C.\ Gu, C.\-D.\ Gong, and D.\ N.\ Sheng,
Phys.\ Rev.\ Lett.\ \textbf{107}, 146803 (2011).


\bibitem{Regnault11}
N. Regnault and B. A. Bernevig,
Phys.\ Rev.\ X \textbf{1}, 021014 (2011).

\bibitem{Venderbos2012}
J.\ Venderbos, S.\ Kourtis, J.\ van den Brink, and
M.\ Daghofer, 
Phys.\ Rev.\ Lett.\ \textbf{108}, 126405 (2012). 

\bibitem{Wu2012}
  Y.-L. Wu, B. Bernevig, and N. Regnault, 
Phys.\ Rev.\ B \textbf{85}, 075116 (2012). 

\bibitem{Qi2012}
  X.-L. Qi, 
Phys.\ Rev.\ Lett.\ \textbf{107}, 126803 (2011).

\bibitem{Parameswaran2012}
S. A. Parameswaran, R. Roy, and S. L. Sondhi, 
Phys.\ Rev.\ B \textbf{85}, 241308 (2012).


\bibitem{Bernevig2012}
  B. A. Bernevig and N. Regnault, 
Phys.\ Rev.\ B \textbf{85}, 075128 (2012).

\bibitem{Wu2012-06}
  Yang-Le Wu, N. Regnault, and B. Andrei Bernevig,
Phys.\ Rev.\ B \textbf{86}, 085129 (2012). 


\bibitem{Wu2012-10}
  Yang-Le Wu, N. Regnault, and B. Andrei Bernevig, 
\texttt{arXiv:1210.6356}. 


\bibitem{LuRan2012}
Yuan-Ming Lu and Ying Ran, 
Phys.\ Rev.\ B \textbf{85}, 165134 (2012).


\bibitem{McGreevy2012}
John McGreevy, Brian Swingle, and Ky-Anh Tran,
Phys.\ Rev.\ B \textbf{85}, 125105 (2012).


\bibitem{Maciejko10}
J.\ Maciejko,
X.\ -L.\ Qi,
A.\ Karch,
and
S. -C. Zhang, 
Phys. Rev. Lett. \textbf{105}, 246809 (2010).

\bibitem{Maciejko11}
J.\ Maciejko,
X.\ -L.\ Qi,
A.\ Karch,
and
S. -C. Zhang, 
Phys.\ Rev.\ B \textbf{86}, 235128 (2012). 

\bibitem{Swingle10}
B. Swingle, M. Barkeshli, J. McGreevy, 
and T. Senthil,
Phys. Rev. B \textbf{83}, 195139 (2011). 



\bibitem{Hansson2011}
T. H. Hansson, A. Karlhede, M. Sato,
\texttt{arXiv:1105.5031},


\bibitem{Fradkin1999}
E. Fradkin, C. Nayak, K. Schoutens, 
Nucl.\ Phys.\ B \textbf{546},  711 (1999).

\bibitem{Fradkin1997}
E. Fradkin, C. Nayak, A. Tsvelik, F. Wilczek, 
Nucl.\ Phys.\ B \textbf{516} 704, (1998).







\end{thebibliography}
\end{document}